\documentclass[%
twocolumn,
superscriptaddress,
 amsmath,amssymb,
 aps,
prr,
]{revtex4-2}

\usepackage{graphicx}
\usepackage{dcolumn}
\usepackage{bm}
\usepackage{color}
\usepackage{ulem}
\usepackage{stackengine}
\usepackage{braket}
\usepackage{multirow}

\usepackage{comment}

\newcommand{\figref}[1]{Fig.~\ref{fig:#1}}
\newcommand{\secref}[1]{Sec.~\ref{sec:#1}}
\newcommand{\equref}[1]{Eq.~(\ref{eq:#1})}

\begin{document}

\title{
Analytical solution of the Schr\"{o}dinger equation with $1/r^3$ and attractive $1/r^2$ potentials: Universal three-body parameter of mixed-dimensional Efimov states}%
\author{Yuki Ohishi}
\thanks{These authors contributed equally to this work.}
\affiliation{Department of Engineering Science, University of Electro-Communications, Chofu, Tokyo 182-8585, Japan}
\author{Kazuki Oi}
\thanks{These authors contributed equally to this work.}
\affiliation{Department of Physics, Tohoku University, Sendai 980-8578, Japan}
\affiliation{Department of Engineering Science, University of Electro-Communications, Chofu, Tokyo 182-8585, Japan}
\author{Shimpei Endo}
\affiliation{Department of Engineering Science, University of Electro-Communications, Chofu, Tokyo 182-8585, Japan}
\affiliation{Institute for Advanced Science, University of Electro-Communications, Chofu, Tokyo 182-8585, Japan}

\date{\today}


\begin{abstract}
We study the Schr\"{o}dinger equation with $1/r^3$ and attractive $1/r^2$ potentials. Using the quantum defect theory, we obtain analytical solutions for both repulsive and attractive $1/r^3$ interactions. The obtained discrete-scale-invariant energies and wave functions, validated by excellent agreement with numerical results, provide a natural framework for describing the universality of Efimov states in mixed dimension. Specifically, we consider a three-body system consisting of two heavy particles with large dipole moments confined to a quasi-one-dimensional geometry and resonantly interacting with an unconfined light particle. With the Born–Oppenheimer approximation, this system is effectively reduced to the Schr\"{o}dinger equation with $1/r^3$ and $1/r^2$ potentials, and manifests the Efimov effect. Our analytical solution suggests that, for repulsive dipole interactions, the three-body parameter of the mixed-dimensional Efimov states is universally set by the dipolar length scale, whereas for attractive interactions it explicitly depends on the short-range phase. We also investigate the effects of finite transverse confinement and find that our analytical results are useful for describing the Efimov states composed of two polar molecules and a light atom.
\end{abstract}

\maketitle


\section{Introduction}

Cold atoms and molecules have recently emerged as versatile platforms for exploring a wide range of quantum phenomena~\cite{RevModPhys.80.885,schafer2020tools,cornish2024quantum}. Owing to their high degree of controllability, cold-atom systems have enabled the realization of various exotic effects that are unique to strongly interacting quantum matters. A prime example of such exotic few-body phenomena is the Efimov effect~\cite{kraemer2006evidence,efimov1970energy,efimov1973energy,RevModPhys.89.035006,naidon2017efimov,d2018few}: a series of three-body bound states exhibiting discrete scale invariance emerges when the $s$-wave scattering lengths characterizing the strength of short-range interactions are resonantly enhanced via a Feshbach resonance~\cite{inouye1998observation,chin2010feshbach}. As the Efimov states can appear in various physical systems such as $^4$He clusters~\cite{PhysRevLett.95.063002,kunitski2015observation}, condensed matter~\cite{nishida2013efimov}, to nuclear physics~\cite{AnnRev_HamPlatt,endoepelbaumQcl2024,PhysRevLett.100.192502,higa2008alphaalpha,PhysRevLett.120.052502,endotanaka2023}, they have been explored in cold-atom experiments to unveil their universal nature~\cite{kraemer2006evidence,Zaccanti2009,PhysRevLett.103.163202,doi:10.1126/science.1182840,PhysRevLett.113.240402,PhysRevLett.112.250404}.

More recently, cold atoms and molecules with a long-range $1/r^3$ interaction have been established as platforms for exploring even more exotic quantum phenomena~\cite{Moses2017,Chomaz_2023}, such as the supersolid phase~\cite{PhysRevX.9.011051,PhysRevLett.122.130405,PhysRevX.9.021012}, Einstein de Haas effect~\cite{PhysRevLett.96.080405,matsui2025observation}, quantum magnetism~\cite{micheli2006toolbox,science.adq0911,li2023tunable}, and controlled chemical reactions~\cite{ni2010dipolar,ospelkaus2010quantum,PhysRevLett.116.205303,karman2024ultracold}. 
In addition to serving as platforms for quantum many-body simulations, polar molecules - owing to their long-range interactions and the suppression of inelastic losses via microwave shielding~\cite{matsuda2020resonant,anderegg2021observation} - have also been recognized as promising candidates for quantum computation~\cite{PhysRevLett.88.067901,picard2025entanglement,cornish2024quantum}. To obtain a detailed account of the exotic quantum phenomena induced by long-range dipolar interactions and to achieve quantum computation with molecules, a microscopic understanding of the underlying two- and three-body processes is crucial.

Because the dipole–dipole interaction scales as $1/r^{3}$ at large distance, its low-energy few-body scattering properties differ markedly from those of short-range interactions~\cite{PhysRev.73.1002,landauQM}. For the two-body system, the effective range expansion, which is widely used in cold-atom studies, is not applicable owing to the peculiar threshold behavior, giving rise to a distinct class of universality characterized by the dipole length scale~\cite{bohn2009quasi,PhysRevA.64.022717,PhysRevA.78.040703}. This anomalous two-body scattering property strongly influences three- and many-body physics. Indeed, the Efimov states with dipolar interactions have been demonstrated to depend universally on the dipole length scale~\cite{PhysRevLett.106.233201,PhysRevLett.107.233201}, in stark contrast to systems of shorter-range $1/r^6$ van der Waals interactions, where the Efimov states are universally determined by the van der Waals length~\cite{PhysRevLett.107.120401,gross2011study,pascaleno3BP1,pascaleno3BP2,PhysRevLett.108.263001,PhysRevLett.109.243201,PhysRevA.95.062708}.

To obtain a quantitatively better understanding of few- and many-body quantum systems based on accurate two-body correlations, analytical solutions of the Schr\"{o}dinger equation with $1/r^n$ potentials are highly valuable. After Gao analytically solved the Schr\"{o}dinger equation with an attractive $1/r^6$ potential using the quantum defect theory and its low-energy expansion~\cite{PhysRevA.58.1728,PhysRevA.64.010701,BoGao2004}, a series of analytical solutions has been derived for various classes of $1/r^n$ interactions: $1/r^3$ potential~\cite{PhysRevA.78.012702,PhysRevA.59.2778,PhysRevA.64.022717}, general $1/r^n$ ($n>2$) potentials~\cite{PhysRevA.78.012702}, repulsive $1/r^6$ potential~\cite{Jie_2025}, harmonically trapped system~\cite{du2024analytical}, an anisotropic $1/r^3$ potential~\cite{PhysRevResearch.7.023162}, and multi-channel systems~\cite{PhysRevA.72.042719,PhysRevA.84.042703}. The method has further been extended to systems involving both attractive $1/r^6$ and $1/r^2$ potentials~\cite{OiEndo2024} to investigate the Efimov physics, revealing the van der Waals universality of Efimov states by deriving analytical expressions for their binding energies and wave functions.

Here, we obtain analytical solutions of the Schr\"{o}dinger equation with $1/r^3$ and attractive $1/r^2$ potentials using the quantum defect theory. For a repulsive $1/r^3$ potential, the binding energies and wave functions are independent of short-range physics and are universally determined by the length scale of $1/r^3$ interaction. In contrast, for an attractive $1/r^3$ potential, our analytical expressions explicitly depend on short-range physics through the quantum defect parameter. These analytical solutions are valuable for describing Efimov states appearing in mixed-dimensional geometries~\cite{nishida2011liberating,PhysRevLett.104.153202}. To illustrate this point, we apply our analytical solutions to a three-body system composed of two heavy dipolar particles confined in a tight cigar-shaped trap and interacting with a light particle with a large $s$-wave scattering length, which leads to the Born-Oppenheimer equation of the form of the Schr\"{o}dinger equation with $1/r^3$ and attractive $1/r^2$ potentials. Our analytical solutions provide explicit formulae of the three-body parameters of the mixed-dimensional Efimov states, revealing that the nature of universality differs between repulsive and attractive $1/r^3$ interactions. We show that our analytical solutions can well describe the behavior of polar molecules~\cite{Moses2017, ni2010dipolar, ospelkaus2010quantum, PhysRevLett.116.205303, karman2024ultracold, PhysRevA.73.022707, matsuda2020resonant, anderegg2021observation,shi2025universal} confined in a tight cigar-shaped trap, whereas the influence of the transverse trapping potential is non-negligible in the case of dipolar atoms~\cite{PhysRevLett.95.150406,PhysRevLett.108.210401,PhysRevLett.107.190401}.

This paper is organized as follows. In \secref{model}, we first explain how the $1/r^2$ and $1/r^3$ potentials arise in mixed-dimensional systems of atoms and molecules with dipole-dipole interactions, and then present the analytical solutions of the corresponding Schr\"{o}dinger equation at the unitary limit. In \secref{result}, we compare the analytical solutions with the numerical results and discuss the experimental feasibility of realizing our system, including discussions on the effect of the transverse trapping potential. Finally, we summarize our findings in \secref{Concl}. Throughout this paper, we use natural units with $\hbar=1$.


\section{\label{sec:model}Analytical solution with $1/r^3$ and $1/r^2$ potentials and Efimov physics with dipole interaction in mixed dimension}

In this section, we present the analytical solutions of the Schr\"{o}dinger equation under $1/r^2$ and $1/r^3$ potentials
\begin{align}
 \left[-\frac{1}{2\mu} \frac{d^2}{dr^2} + \frac{s^2-\frac{1}{4}}{2\mu r^2}  -\frac{C_3}{r^3}\right] \psi (r) = E \psi (r),
 \label{eq:onedimSchEq}
\end{align}
where $\mu$ is the reduced mass, and $C_3$ is the strength of repulsive ($C_3<0$) and attractive ($C_3>0$) $1/r^3$ interaction. The characteristic length scale  defined as $\beta_3 = 2\mu |C_3| $ serves as the dipolar analogue of the van der Waals length $\beta_6$ characterizing the van der Waals universality of Efimov states in Ref.~\cite{OiEndo2024}. $s^2$ is a constant characterizing the strength of the $1/r^2$ potential. For integer values of angular momenta $\ell=0,1,2,...$ with $s^2 = (\ell+1/2)^2 >0$, \equref{onedimSchEq} has been solved analytically with the quantum defect theory (QDT)~\cite{PhysRevA.64.022717, PhysRevA.78.012702, PhysRevA.59.2778, PhysRevA.64.010701, BoGao2004, OiEndo2024,PhysRevResearch.7.023162}. As we show in detail in Sections~\ref{sec:repandatt} and~\ref{sec:attandatt}, this equation can also be solved analytically for arbitrary negative values $s^2<0$ (i.e., a pure imaginary $s$).

\begin{figure}[!t]
	\centering
	\includegraphics[width=1.02\linewidth]{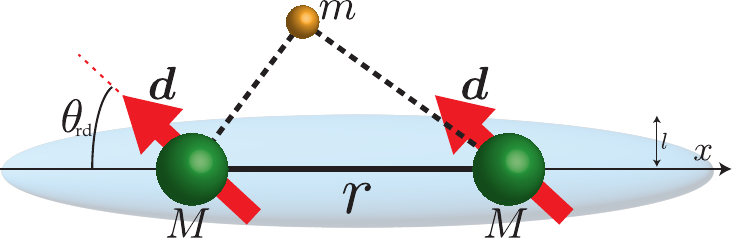} 
\caption{Schematic illustration of the three-body system studied in this paper~: we consider a system two heavy particles (mass $M$) with dipole moments $\bm{d}$ interacting with a light particle (mass $m$) with an $s$-wave scattering length $a^\mathrm{(HL)}$ (black dotted line). The heavy particles are assumed to be confined in an axially symmetric harmonic trap (trapping length $l$), while the light particle moves freely in a three dimensional space. The direction of the dipole moment $\theta_\mathrm{rd}$ can be controlled by an external field.}
\label{fig:SchematicFig}
\end{figure}

Equation~\eqref{eq:onedimSchEq} with $s^2<0$ is relevant to Efimov states of cold atoms and molecules interacting with dipole-dipole interactions in a mixed dimension. To illustrate this point, we consider a three-body system composed of two heavy particles (mass $M$) and a light particle (mass $m$) as illustrated in \figref{SchematicFig}. We consider a mixed dimensional system~\cite{nishida2011liberating,PhysRevLett.104.153202}, such that the heavy particles are confined by an axially symmetric harmonic trap (see \equref{HOtrap}) while the light particle moves freely in a three-dimensional space. The heavy particles are assumed to possess large dipole moments, such that they interact predominantly via a dipole-dipole interaction at large distance. Such a system can be realized with highly magnetic cold atoms, such as Cr~\cite{PhysRevLett.95.150406}, Er~\cite{PhysRevLett.108.210401}, and Dy~\cite{PhysRevLett.107.190401}, or with polar molecule~\cite{Moses2017, ni2010dipolar, ospelkaus2010quantum, PhysRevLett.116.205303, karman2024ultracold, PhysRevA.73.022707, matsuda2020resonant, anderegg2021observation,shi2025universal}. On the other hand, the light particle is assumed to have a negligibly small dipole moment, so that the interaction between the light and heavy particles is of the short-range van der Waals type. We model this short-range interaction between the heavy and light particles by a zero-range contact interaction characterized by the $s$-wave scattering length $a^\mathrm{(HL)}$~\cite{bethe1935scattering}. This assumption is particularly appropriate for describing low-energy bound and scattering states of cold atoms and molecules~\cite{chin2010feshbach}.

When the mass ratio between the heavy and light particle is large $M/m \gg 1$, the system can be well described by the Born-Oppenheimer approximation, where the effect of the light particle is incorporated as an effective interaction between the heavy particles. The Schr\"{o}dinger equation for the relative motion of the two-heavy particles (dipole moments $\bm{d}$) is written as
\begin{equation}
    \left[ -\frac{\nabla^2_{\bm{r}}}{M} + V_\mathrm{BO}(r) + V_\mathrm{dd}(\bm{r}) + V_\mathrm{HO}(\bm{r}) \right] \Psi(\bm{r}) = E\Psi(\bm{r}).
    \label{eq:SchEqVBO}
\end{equation}
Here, $V_\mathrm{HO}$ is an axially symmetric external harmonic potential acting solely on the heavy particles 
\begin{equation}
    V_\mathrm{HO}(\bm{r}) = \frac{M}{4}\omega^2 (y^2+z^2)
    \label{eq:HOtrap}
\end{equation}
with the transverse trapping frequencies $\omega$, and $V_\mathrm{dd}$ is the dipole-dipole interaction between the heavy particles
\begin{equation}
    V_\mathrm{dd}(\bm{r}) = \frac{C_\mathrm{dd}(1-3~\mathrm{cos}^2\theta_\mathrm{rd})}{r^3},
\end{equation}
where $C_\mathrm{dd}$ denotes the dipole-dipole coupling constant and $\theta_{\mathrm{rd}}$ is the angle between $\bm{r}$ and $\bm{d}$. $V_\mathrm{BO}(r)$ is the Born-Oppenheimer potential originating from the exchange of the light particle, which is analytically obtained as~\cite{FONSECA1979273}
\begin{equation}
    V_\mathrm{BO}(r) = -\frac{1}{2mr^2} \left( \frac{r}{a^\mathrm{(HL)}} + W \left( e^{-\frac{r}{a^\mathrm{(HL)}}} \right) \right)^2 \Theta \left( \frac{r}{a^\mathrm{(HL)}} +1 \right),
    \label{eq:BOPotential}
\end{equation}
where $W$ and $\Theta$ are the Lambert W function and the Heaviside step function, respectively.

When the transverse trapping frequency is large, the heavy particles can be considered to remain in the transverse ground state. For such a tight cigar-shaped trap, the wave function is approximated as $\Psi(\bm{r})=\psi(x)\varphi_0(y,z)$ where $\psi(x)$ is the wave function along the axial direction $x$, and $\varphi_0$ is the ground-state wave function of the transverse harmonic potential. Integrating $V_\mathrm{dd}(\bm{r})$ over the transverse directions ($y$ and $z$) and denoting $x$ as $r$, we obtain a one-dimensional Schr\"{o}dinger equation
\begin{equation}
    \left[ -\frac{1}{M} \frac{d^2}{dr^2} + V_\mathrm{BO}(r) + V_\mathrm{q1D}(r) \right] \psi(r) = E\psi(r)
    \label{eq:q1Dscheq}
\end{equation}
with an effective quasi-one-dimensional dipole-dipole interaction~\cite{PhysRevLett.99.140406, PhysRevA.81.063616}
\begin{equation}
    V_\mathrm{q1D}(r) = -\frac{C_\mathrm{dd} [1+3~\mathrm{cos}2\theta]}{8 l^3} v(r/l),
    \label{eq:q1deq_vpotang}
\end{equation}
with
\begin{equation}
    v(x) = -2x + \sqrt{2\pi}(1+x^2)e^\frac{x^2}{2} \mathrm{erfc}(x/\sqrt{2}).
\end{equation}
Here, $l = \sqrt{\hbar / M\omega}$ is the harmonic oscillator length of the transverse confinement, and $\theta = \theta_{\mathrm{rd}}$ is the angle between the axial direction and $\bm{d}$.

In the unitary limit $a^\mathrm{(HL)} \rightarrow \pm \infty$, the Born-Oppenheimer potential reduces to $V_\mathrm{BO}(r) \sim 1/r^2$. Furthermore, considering the limit of a tight cigar-shaped trap $\omega \rightarrow \infty$, the effective potential reduces to $V_\mathrm{q1D} (r) \sim 1/r^3$. We thus obtain \equref{onedimSchEq} with $\mu =M/2$ and 
\begin{equation}
\label{eq:sell_def}s^2 = \frac{1}{4} - \frac{M\Omega^2}{2m},
\end{equation}
where $\Omega = W(1)=0.5671...$, and 
\begin{equation}
\label{eq:C3_theta_rel}
C_3 = C_{\mathrm{dd}}(1+ 3 \cos2\theta)/2.
\end{equation}
Due to the $1/r^2$ attraction ($s^2<0$), it shows the Efimov effect~\cite{nishida2011liberating}. The analytical solutions of \equref{onedimSchEq}, derived in \secref{repandatt} and~\ref{sec:attandatt}, are thus relevant for the Efimov physics of dipolar atoms or polar molecules in a mixed dimension. Such a system can be experimentally realized by a mixture of light cold atoms with heavy cold dipolar atoms or polar molecules selectively confined in an optical lattice, with the heavy-light interaction fine-tuned as $a^\mathrm{(HL)} \rightarrow \pm \infty$ by the Feshbach resonance~\cite{ErLiFR1,ErLiFR2,DyLifeshbach,CrLiFR,yang2022creation, doi:10.1126/science.aau5322, chen2023field}. Notably, by varying the direction of the external magnetic or electric fields, the direction of the dipole moment $\bm{d}$ hence the strength of $1/r^3$ interaction $C_3$ can be controlled to realize either repulsive $C_3<0$ or attractive $C_3>0$ interactions~\cite{ni2010dipolar,ospelkaus2010quantum,PhysRevLett.95.150406,PhysRevLett.108.210401,PhysRevLett.107.190401, doi:10.1126/sciadv.adr2715}.


\subsection{\label{sec:repandatt}Repulsive $1/r^3$ and attractive $1/r^2$ potentials }
The analytical solutions of \equref{onedimSchEq} can be obtained in a similar manner to that employed in Ref.~\cite{OiEndo2024}: By generalizing Gao’s quantum defect theory (QDT)~\cite{PhysRevA.59.2778} to the case of $s^2 <0$, a general solution of \equref{onedimSchEq} is expressed as a linear combination of the two independent special solutions $f^c(r)$ and $g^c(r)$ as (see \equref{repulsive_AMI_solution_app} in the Appendix for concrete expressions of $f^c$ and $g^c$)
\begin{equation}
    \psi(r) = \mathcal{N} [f^c(r) - K^c g^c(r)],
    \label{eq:QDTWF_fcKcgc}
\end{equation}
where $K^c$ is the QDT parameter which characterizes the phase shift at short-range region $r\ll \beta_3$ and $\mathcal{N}$ is a normalization constant. 

For bound states, the solution must satisfy the boundary condition $\psi(r\rightarrow\infty)=0$, from which we obtain
\begin{equation}
\label{eq:BCforBoundstate_repulsive}
    K^c=\frac{W_{f-}}{W_{g-}}=\dfrac{1}{\sin[2\pi(\nu-\nu_0)]}\dfrac{1-\mathcal{M}}{1+\mathcal{M}},
\end{equation}
where $\nu_0=s = i |s|$, $W_{f-}$ and $W_{g-}$ are given as Eqs.~(32) and (34) in Ref.~\cite{PhysRevA.59.2778}, and $\mathcal{M}$ defined as in \equref{M_def} can be evaluated at low energy $|\Delta| \ll 1$ (with $\Delta\equiv \sqrt{2\mu E}\beta_3/2$) as

\begin{equation}
    \begin{split}
        \mathcal{M} &\approx |\Delta|^{2\nu_0} \frac{\Gamma(1-2\nu_0)}{\Gamma(1+\nu_0)} \frac{\Gamma(1-\nu_0)}{\Gamma(1+2\nu_0)}.
        \label{eq:M_low_ene_exp}
    \end{split}
\end{equation}
In the absence of the Efimov effect, $\nu_0$ is real, and thus $\mathcal{M} \simeq 0$ in the low-energy limit $|\Delta| \rightarrow 0$. Indeed, the binding energy and low-energy scattering parameters for a two-body problem with a $1/r^3$ potential in the $\ell$-th angular momentum channel can be obtained from Eqs.~(\ref{eq:BCforBoundstate_repulsive}) and ~(\ref{eq:M_low_ene_exp}) by setting $\nu_0=2\ell+1$~\cite{PhysRevA.59.2778,PhysRevA.64.022717}. In contrast, when the system has an attractive $1/r^2$ potential with $s^2 < 0$ as in \equref{onedimSchEq}, $\nu_0$ becomes pure imaginary $\nu_0= i|s|$. Consequently, the behavior of $\mathcal{M}$ at low energy $|\Delta| \rightarrow 0$ is modified as
\begin{equation}
        \mathcal{M} \cong -e^{i \{ 2|s| \ln|\Delta| - 2\xi \}},
    \label{eq:Mel_exponential}
\end{equation}
where 
\begin{equation}
\xi \equiv \mathrm{arg} [\Gamma(i|s|)\Gamma(1+2i|s|)].
\end{equation}
The log-periodic behavior of \equref{Mel_exponential} as a function of energy $\Delta =\sqrt{2\mu E}\beta_3/2$ is a clear manifestation of the discrete-scale invariance of the Efimov effect.

As shown in the Appendix (see \equref{ShortRangeRepulsive}), $f^c$ decreases exponentially as $r\rightarrow0$ while $g^c$ increases exponentially, corresponding to a regular and irregular solution, respectively. If the length scale associated with the short-range physics $R_{\mathrm{min}}$ is small $R_{\mathrm{min}} \ll \beta_3$, as is the case for cold atoms and molecules, the wave function should be solely described by the regular solution $f^c$. In other words, the wave function in the short-range region $r \ll \beta_3$ should be suppressed due to the repulsive $1/r^3$ potential, which results in $K^c\cong0$, or equivalently, $\mathcal{M}\cong1$. 
Substituting this into \equref{Mel_exponential}, we obtain an analytical expression of the three-body binding energy as
\begin{equation}
    E_n = -\frac{2}{\mu\beta_3^{~2}} \mathrm{exp} \left( \frac{\pi}{|s|} + \frac{2}{|s|} \xi \right) \times e^{-\frac{2n\pi}{|s|}},
    \label{eq:EnergyRepDipole}
\end{equation}
where $n$ is an integer quantum number characterizing the $n$-th Efimov states. The discrete-scale invariance with a scale factor $\displaystyle e^{2\pi/|s|}$ naturally emerges as an analytical continuation: by converting a real value $\nu_0=2\ell+1$ corresponding to a centrifugal repulsion into a pure imaginary value $\nu_0=s = i |s|$ corresponding an Efimov attraction, the low-energy behavior of $\mathcal{M}$ in \equref{M_low_ene_exp} leads to the discrete-scale-invariant energy spectrum. Equation~(\ref{eq:EnergyRepDipole}) can therefore be regarded as an analytical continuation of the binding-energy formula for a two-body problem with a $1/r^3$ repulsion and a centrifugal barrier obtained in Refs.~\cite{PhysRevA.59.2778,PhysRevA.64.022717}.

Notably, the binding energy in \equref{EnergyRepDipole} is independent of the QDT parameter $K^c$ and the short-range boundary condition $R_{\mathrm{min}}$. In other words, the three-body parameter of the Efimov states is universally determined by $\beta_3$
 characterizing the long-range dipole potential $1/r^3$. This universality arises from the suppression of the wave function at short distance caused by the $1/r^3$ repulsion. It is analogous to the universality found for three identical dipolar bosons in Ref.~\cite{PhysRevLett.106.233201}. While a hyperspherical repulsive potential arising from the coupling of multiple angular-momentum channels is required to describe the three identical dipoles~\cite{PhysRevLett.106.233201}, our mass-imbalanced mixed-dimensional system allows a simple analytical description of the dipolar Efimov universality, which shares common underlying mechanism — a repulsive potential at the scale $\simeq \beta_3$ — as the physical origin of the universality. As another notable analogous system, for three identical bosons interacting via van der Waals forces, the three-body parameter has been found both experimentally~\cite{PhysRevLett.107.120401,gross2011study,PhysRevLett.111.053202,PhysRevLett.123.233402,PhysRevLett.125.243401,johansen2017testing} and theoretically~\cite{PhysRevLett.108.263001,pascaleno3BP1,pascaleno3BP2,PhysRevA.103.052805,PhysRevLett.93.143201,PhysRevLett.100.140404,PhysRevA.86.052516,PhysRevA.107.023301,PhysRevLett.132.133402} to be universally determined by the van der Waals length for broad Feshbach resonances. In this system, the non-adiabatic repulsion appearing universally at the van der Waals scale is demonstrated to suppress short-range configurations, leading to a universal three-body parameter~\cite{PhysRevLett.108.263001,pascaleno3BP1,pascaleno3BP2}.

While the solutions $f^c$, $g^c$ and hence the bound-state wave function at finite binding energy is represented as a complicated Bessel series expansion (see Appendix A and B),  they can be expressed simpler at low energy (or equivalently, at short distance) $r \ll 1/\sqrt{2\mu |E|}$ as
\begin{equation}
    \begin{split}
        f^c(r) &= -\frac{2}{\sinh(2 \pi |s|)} \mathrm{Im}\left[\sqrt{r} I_{2i|s|}\left(2 \sqrt{\frac{\beta_3}{r}} \right) \right],\\
        g^c(r) &= - 2 \mathrm{Re}\left[\sqrt{r} I_{2i|s|}\left(2 \sqrt{\frac{\beta_3}{r}} \right) \right].
        \label{eq:fcgc_shortI_rep}
    \end{split}
\end{equation}
The regular and irregular behavior of $f^c,g^c$ at short distance [\equref{ShortRangeRepulsive})] can also be understood from these expressions, as an asymptotic behavior of the modified Bessel function $I$ with a pure imaginary index.


\subsection{\label{sec:attandatt}Attractive $1/r^3$ and attractive $1/r^2$ potentials }
The analytical solution for an attractive potential $C_3>0$ is obtained with a procedure similar to that in the repulsive case. The two independent special solutions $f^c(r)$ and $g^c(r)$ are given as \equref{attractive_AMI_solution_app}, which is similar but slightly different from the repulsive case in \equref{repulsive_AMI_solution_app}. Using these two solutions, a general solution is expressed by the same formula as \equref{QDTWF_fcKcgc}. The bound-state condition is obtained by imposing the boundary condition $\psi(r\rightarrow\infty)=0$ as
\begin{equation}
    \label{eq:BCforBoundstate_attractive}
    K^c=\frac{W_{f-}}{W_{g-}}=\tan(\nu \pi)\frac{1+\mathcal{M}}{1-\mathcal{M}},
\end{equation}
where $W_{f-}$and $W_{g-}$ are given as \equref{Wfg}, and the definition of $\mathcal{M}$ is same as in the repulsive case.

One notable difference from the repulsive case is that both $f^c$ and $g^c$ are of the same order at short distance and contribute equally  to the wave function, allowing $K^c$ to take an arbitrary value. In other words, the binding energy can explicitly depend on $K^c$, hence short-range phase. To be more specific, by using the low-energy expansion formula \equref{Mel_exponential} and $\nu \simeq s = i |s|$, \equref{BCforBoundstate_attractive} is written as
\begin{equation}
\label{eq:KcAndWfmWgmLowenergyAttractive}
    K^c \cong \tanh(\pi |s|)\tan \left[\frac{|s|}{2}\ln|\Delta|-\xi\right],
\end{equation}
from which the three-body binding energy is obtained as 
\begin{equation}
    E_n = -\frac{2}{\mu \beta_3^{~2}} \mathrm{exp} \left( \frac{2}{|s|} \left[\mathrm{arctan}\left( \frac{K^c}{\mathrm{tanh}(\pi |s|)} \right) +\xi \right] \right) \times e^{-\frac{2n\pi}{|s|}}.
    \label{eq:EnergyAttDipole}
\end{equation}
The discrete-scale invariance of the Efimov states thus naturally emerges as an analytical continuation of the two-body solution under an attractive $1/r^3$ potential and a centrifugal barrier~\cite{PhysRevA.64.022717,PhysRevA.78.012702}. In contrast to the repulsive $1/r^3$ case in \equref{EnergyRepDipole}, the three-body binding energy explicitly depends on the QDT parameter $K^c$. This difference can be ascribed to the short-range behavior of the wave function: Due to the attractive potential $1/r^3$, the wave function can penetrate into the short-range region $r\ll \beta_3$, causing the binding energy to depend explicitly on the short-range detail through $K^c$. To further elucidate this point, we note that at short distance  $r \ll 1/\sqrt{2\mu |E|}$, the solutions $f^c,g^c$ have simpler analytical expressions~\cite{PhysRevA.78.012702} (see Appendix A and B for the derivation), 
\begin{equation}
    \begin{split}
    f^c(r) = \frac{1}{\cosh(\pi |s|)} \mathrm{Re}\left[\sqrt{r} J_{2i|s|}\left(2 \sqrt{\frac{\beta_3}{r}} \right) \right],\\
    g^c(r) = -\frac{1}{\sinh(\pi |s|)} \mathrm{Im}\left[\sqrt{r} J_{2i|s|}\left(2 \sqrt{\frac{\beta_3}{r}} \right) \right].
    \end{split}
    \label{eq:zeroenergy_wavefunction_attractive}
\end{equation}
In contrast to the repulsive case in \equref{fcgc_shortI_rep}, both $f^c$ and $g^c$ oscillate rapidly at short distance and their magnitudes are of the same order [see \equref{ShortRangeAttractive})]. Substituting \equref{zeroenergy_wavefunction_attractive} into the hard-wall boundary condition $\psi(r=R_{\mathrm{min}})=0$, the QDT parameter $K^c$ is related to $R_{\mathrm{min}}$ as 
\begin{align}
        K^c &= -\mathrm{tanh} (\pi |s|) \frac{\mathrm{Re}\left[ J_{2i|s|} \left(2 \sqrt{\frac{\beta_3}{R_{\mathrm{min}}}}\right)\right]}{\mathrm{Im}\left[J_{2i|s|}\left(2 \sqrt{\frac{\beta_3}{R_{\mathrm{min}}}}\right)\right]} \label{eq:kc}\\
        &\cong -\frac{1}{\mathrm{tan}\left[2\sqrt{\frac{\beta_3}{R_\mathrm{min}}} - \frac{\pi}{4}\right]} \label{eq:kc_asympform}.
\end{align}
In the last line, we use the asymptotic form of the Bessel function. This equation demonstrates that the QDT parameter $K^c$ depends on the short-range physics through $R_{\mathrm{min}}$ which incorporates the phase shift acquired by a complicated short-range potential not captured by $V_{\mathrm{dd}}(\bm{r})$.

While the explicit dependence of the binding energy in \equref{EnergyAttDipole} on $K^c$ appears to indicate a nonuniversal sensitivity of the Efimov state to short-range details of the atoms, such as electronic structures and internal spins, we stress that it may still be cast in a universal description by employing a low-energy scattering parameter of the two heavy particles. To elucidate this point, let us neglect the presence of the light particle and consider two heavy particles in the same confined geometry, whose Schr\"{o}dinger equation is given by \equref{onedimSchEq} with $s=1/2$. Although this two-body Schr\"{o}dinger equation differs from the three-body Efimovian system of \equref{onedimSchEq} with $s^2<0$ for $r \gtrsim \beta_3$, the short-range behavior at $r \ll \beta_3$ are the same, as reflected in the $s$-independence in \equref{kc_asympform} (see also the leading order term of \equref{ShortRangeAttractive} in the Appendix). Since the low-energy two-body scattering parameter such as the scattering length is universally related to $K^c$~\cite{PhysRevA.64.022717,PhysRevA.78.012702}, \equref{EnergyAttDipole} can be rewritten as its universal function. This universality is analogous to the van der Waals universality in highly mass-imbalanced systems, where the universal relation between $K^c$ and the $s$-wave scattering length $a^{(\mathrm{HH})}$ of the heavy particles result in a universal three-body parameter determined by the van der Waals length and $a^{(\mathrm{HH})}$~\cite{OiEndo2024,oi2025universal,PhysRevLett.109.243201,PhysRevA.95.062708}. From this perspective, the $K^c$-independent result for the repulsive $1/r^3$ potential in \equref{EnergyRepDipole} can be alternatively interpreted as a case where the low-energy scattering parameter and $K^c$ are universally determined by the dipole length $\beta_3$~\cite{bohn2009quasi,PhysRevA.64.022717,PhysRevA.78.040703}, so that the Efimov states are universally determined by $\beta_3$. In the above argument, as well as in Ref.~\cite{OiEndo2024}, it is implicitly assumed that the short-range three-body phase is dominated by the short-range interaction between the heavy particles, and the contribution from the light particle can be neglected. This assumption is particularly valid for highly mass-imbalanced heavy-heavy-light atomic or molecular systems, where the polarizability of the light particle is relatively small and hence its short-range van der Waals and higher-order forces are correspondingly weaker than those of the heavy particles.


\section{\label{sec:result}Comparison with numerical results}

In this section, we present our numerical solutions of Eqs.~\eqref{eq:onedimSchEq} and \eqref{eq:q1Dscheq}, and compare them with the analytical solutions derived in \secref{model}. Equations~\eqref{eq:onedimSchEq} and \eqref{eq:q1Dscheq} are solved numerically with the finite-difference method by imposing the fixed boundary conditions at relatively short and sufficiently large distances: $\psi(r=R_{\mathrm{min}})=0$ and $\psi(r=R_{\mathrm{max}})=0$ with $R_{\mathrm{min}}\ll \beta_3 \ll R_{\mathrm{max}}$. More specifically, in our finite-difference method, we discretize the radial coordinate between $R_{\mathrm{min}}$ and $R_{\mathrm{max}}$ using an exponential grid. The calculations are performed with a grid of $5000$-$10000$ points ranging with $R_\mathrm{max} / R_\mathrm{min} = 1000$ - $2000$. With these parameters, the numerical relative error of the bound-state energies is confirmed to be of order $10^{-5}$-$10^{-6}$. Given our specific interest to dipolar atoms and polar molecules, we perform the calculations for the mass ratios corresponding to $^{167}$Er-$^{6}$Li atoms ($M/m = 27.7520...$), and $^{6}$Li atom and $^{23}$Na$^{40}$K molecule ($M/m = 10.4659...$), respectively.  

\subsection{One-dimensional limit}

\begin{figure}[!t]
	\centering
	\includegraphics[width=1.0\linewidth]{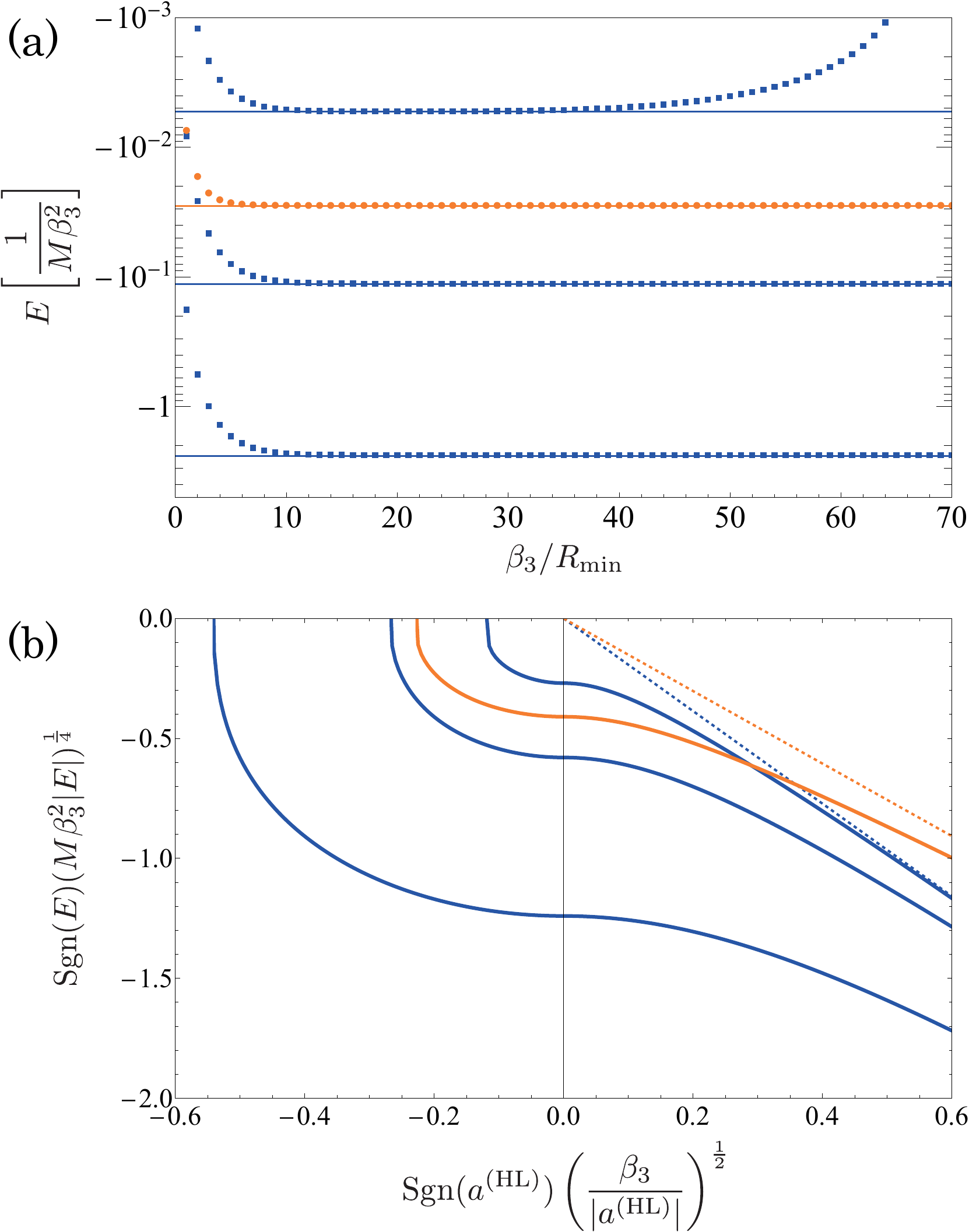} 
\caption{(a) Energy spectra of the Efimov bound states under a repulsive $1/r^3$ potential  at the unitary limit $1/a^{(\mathrm{HL})}=0$ calculated with $R_\mathrm{max}/R_\mathrm{min} = 2000$. Blue squares (orange circles) show the energies obtained by numerically solving \equref{onedimSchEq} for mass ratios corresponding to $^{167}$Er-$^{6}$Li atoms $M/m = 27.7520...$ ($^{6}$Li atom and $^{23}$Na$^{40}$K molecule $M/m = 10.4659...$), respectively. Solid curves denote the analytical result in Eq.~(\ref{eq:EnergyRepDipole}) with $\mu=M/2$. (b) Energy spectra for finite heavy-light $s$-wave scattering length $a^{(\mathrm{HL})}$, calculated with $\beta_3 / R_\mathrm{min} = 16.0$. We only show three and one states among the whole Efimov series for $^{167}$Er-$^{6}$Li (blue) and $^{23}$Na$^{40}$K-$^{6}$Li (orange), respectively. Dashed lines denote the heavy-light dimer states, into which the Efimov trimers dissociate for $a^{(\mathrm{HL})}\rightarrow +0$. }
\label{fig:repulsive_1DUnitary_BE}
\end{figure}

\begin{figure}[!t]
	\centering
	\includegraphics[width=1.0\linewidth]{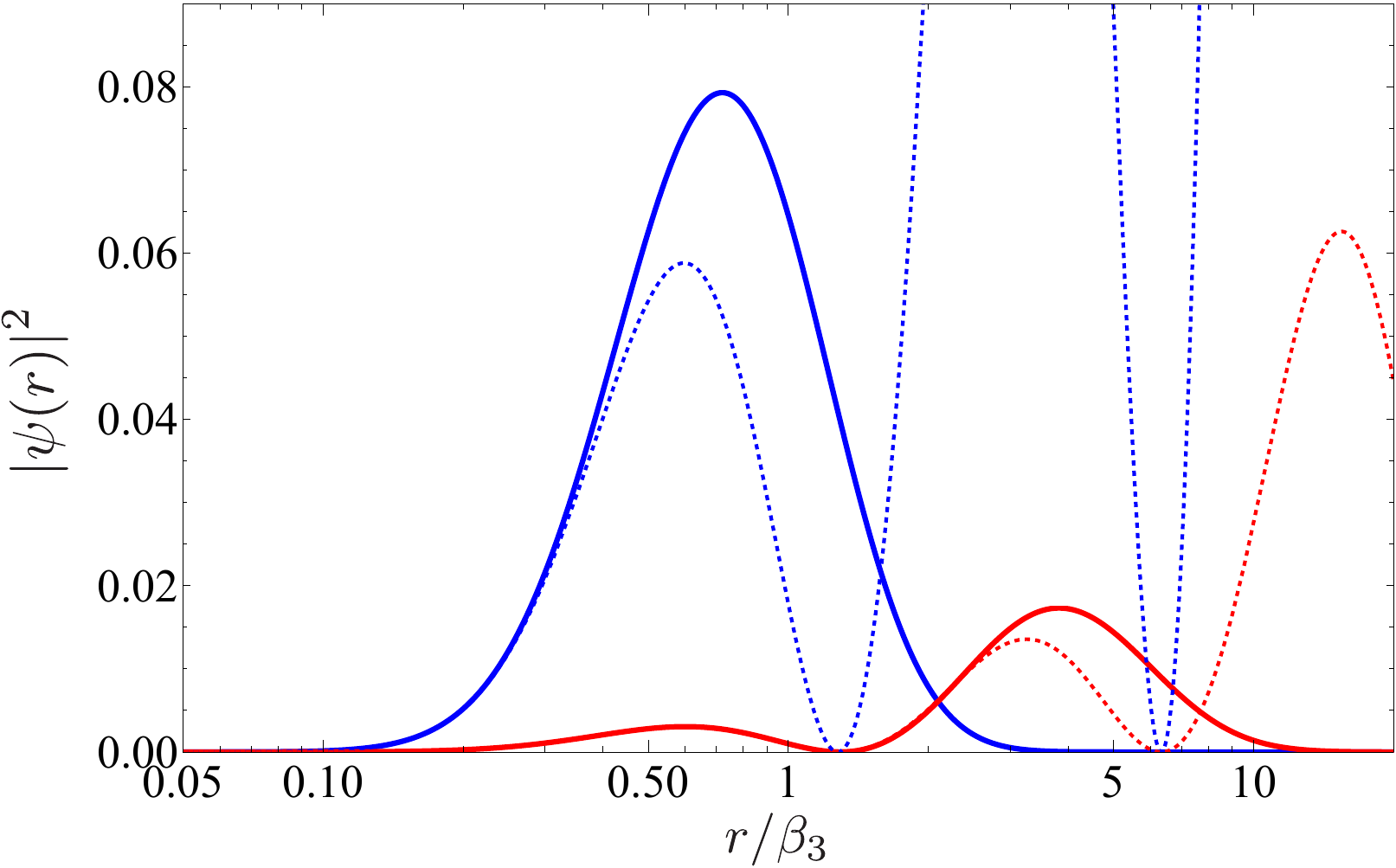} 
\caption{Efimov states' wave function for a repulsive $1/r^3$ potential, obtained by solving \equref{onedimSchEq} with $M/m = 27.7520...$, $R_\mathrm{min} / \beta_3 = 0.05$, and $R_\mathrm{max} / \beta_3 = 20.0$. Blue (red) solid curves are the ground and first-excited Efimov states, respectively. Dashed curves denote analytical solutions for $r \ll 1/\sqrt{M |E|}$ in Eqs.~(\ref{eq:fcgc_shortI_rep}) and (\ref{eq:QDTWF_fcKcgc}) with $K^c= f^c (R_{\mathrm{min}})/g^c (R_{\mathrm{min}})$.}
\label{fig:RepulWavefunc}
\end{figure}

In \figref{repulsive_1DUnitary_BE}(a), we show the numerical solution of \equref{onedimSchEq} with a repulsive $1/r^3$ potential $C_3<0$ at the unitary limit $1/a^{(\mathrm{HL})}=0$. The discrete-scale-invariant energy spectra obtained numerically agree excellently with the analytical expression (solid lines) in \equref{EnergyRepDipole} with $\mu =M/2$ for $10 \lesssim \beta_3/R_{\mathrm{min}} \lesssim 40$. This demonstrates that the three-body parameter of the Efimov states is insensitive to $R_{\mathrm{min}}$ and is instead universally determined by the dipole interaction length scale $\beta_3$, due to a strong repulsive potential that prevents the particles from getting close. This feature is also evident in the behavior of the wave function shown in \figref{RepulWavefunc}: the wave function is exponentially suppressed at short distances $r \ll \beta_3$ for both ground (blue) and first-excited (red) Efimov states. The numerical results (solid curves) are in agreement with the analytical expression for $r \ll 1/\sqrt{M|E|}$ in \equref{fcgc_shortI_rep} (dashed curves) [see also the leading order term of \equref{ShortRangeRepulsive}], whereas they deviate at long distances where the condition $r \ll 1/\sqrt{M|E|}$ to derive \equref{fcgc_shortI_rep} is not satisfied. 

For a loosely bound state of $M \beta_3^2 |E| \lesssim 10^{-2}$ and $\beta_3/R_{\mathrm{min}} \gtrsim 50$ [upper-right region in \figref{repulsive_1DUnitary_BE}(a)], the finite-size effect in our numerical calculation of $R_\mathrm{max}/R_\mathrm{min}=2000$ is visible as a discrepancy between the analytical and numerical results because the spatial extent of the bound state becomes comparable to the system size $R_\mathrm{max}$. On the other hand, the discrepancy for $\beta_3/R_{\mathrm{min}} \lesssim 5$ in \figref{repulsive_1DUnitary_BE}(a) originates from the breakdown of the condition $\beta_3/R_{\mathrm{min}}\gg 1$ used in deriving \equref{EnergyRepDipole}. Indeed, in the limit of $\beta_3/R_{\mathrm{min}}\rightarrow 0$, the binding energy of the Efimov states becomes $E \propto 1/M R_{\mathrm{min}}^2$, which is distinct from $E \propto 1/M \beta_3^2$ of \equref{EnergyRepDipole}.

The energy spectra away from the unitary limit $1/a^{(\mathrm{HL})}\neq 0$ are shown in \figref{repulsive_1DUnitary_BE}(b). The discrete-invariant spectra of the Efimov states are clearly demonstrated, together with the Borromean character for $a^{(\mathrm{HL})}<0$ and particle-dimer dissociation for $a^{(\mathrm{HL})}>0$~\cite{efimov1970energy,efimov1973energy,RevModPhys.89.035006,naidon2017efimov,d2018few}. We confirm that the results obtained for different values of $\beta_3/R_{\mathrm{min}}$ collapse excellently onto the same curves, demonstrating the universality, with a small noticeable deviation for the most tightly bound Efimov state in \figref{repulsive_1DUnitary_BE}(b). Although we also find deeply bound trimers of $M \beta_3^2|E| \gtrsim 100$ (not shown in \figref{repulsive_1DUnitary_BE}(a) and (b)), these states do not exhibit discrete-scale invariance or universality. We therefore identify the trimer with $M \beta_3^2|E| \sim  1$ as the ground Efimov trimer.

\begin{figure}[!t]
	\centering
	\includegraphics[width=1.0\linewidth]{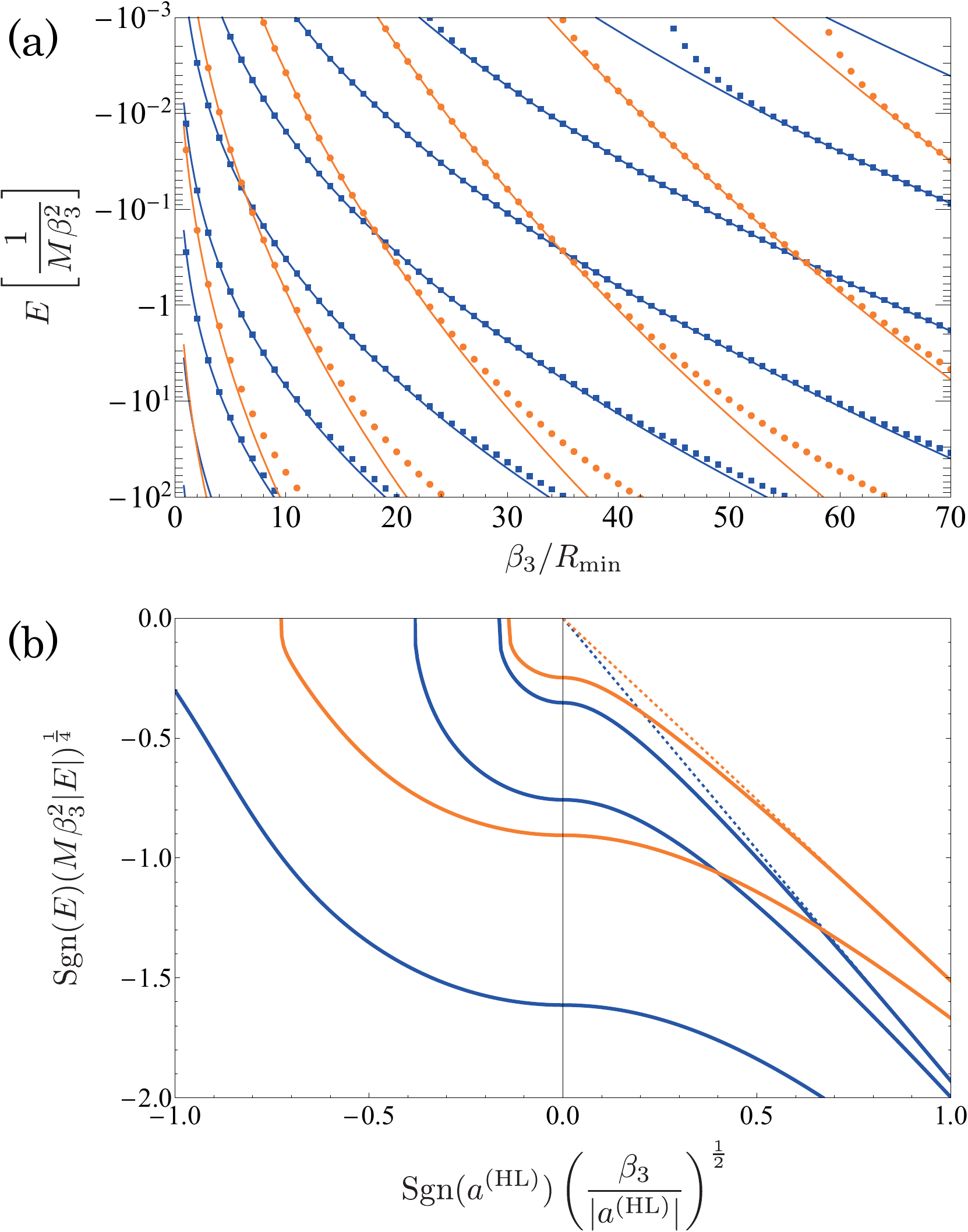} 
\caption{(a) Energy spectra of the Efimov bound states under an attractive $1/r^3$ potential at the unitary limit $1/a^{(\mathrm{HL})}=0$. Solid curves denote the analytical result in Eqs.~(\ref{eq:EnergyAttDipole}) and (\ref{eq:kc}) with $\mu=M/2$. (b) Energy spectra for finite heavy-light $s$-wave scattering length $a^{(\mathrm{HL})}$, obtained with 
$\beta_3 / R_\mathrm{min} = 10.0$. We only show three and two states among the whole Efimov series for $^{167}$Er-$^{6}$Li (blue) and $^{23}$Na$^{40}$K-$^{6}$Li (orange), respectively.}
\label{fig:attractive_1DUnitary_BE}
\end{figure}

\begin{figure}[!t]
	\centering
	\includegraphics[width=1.0\linewidth]{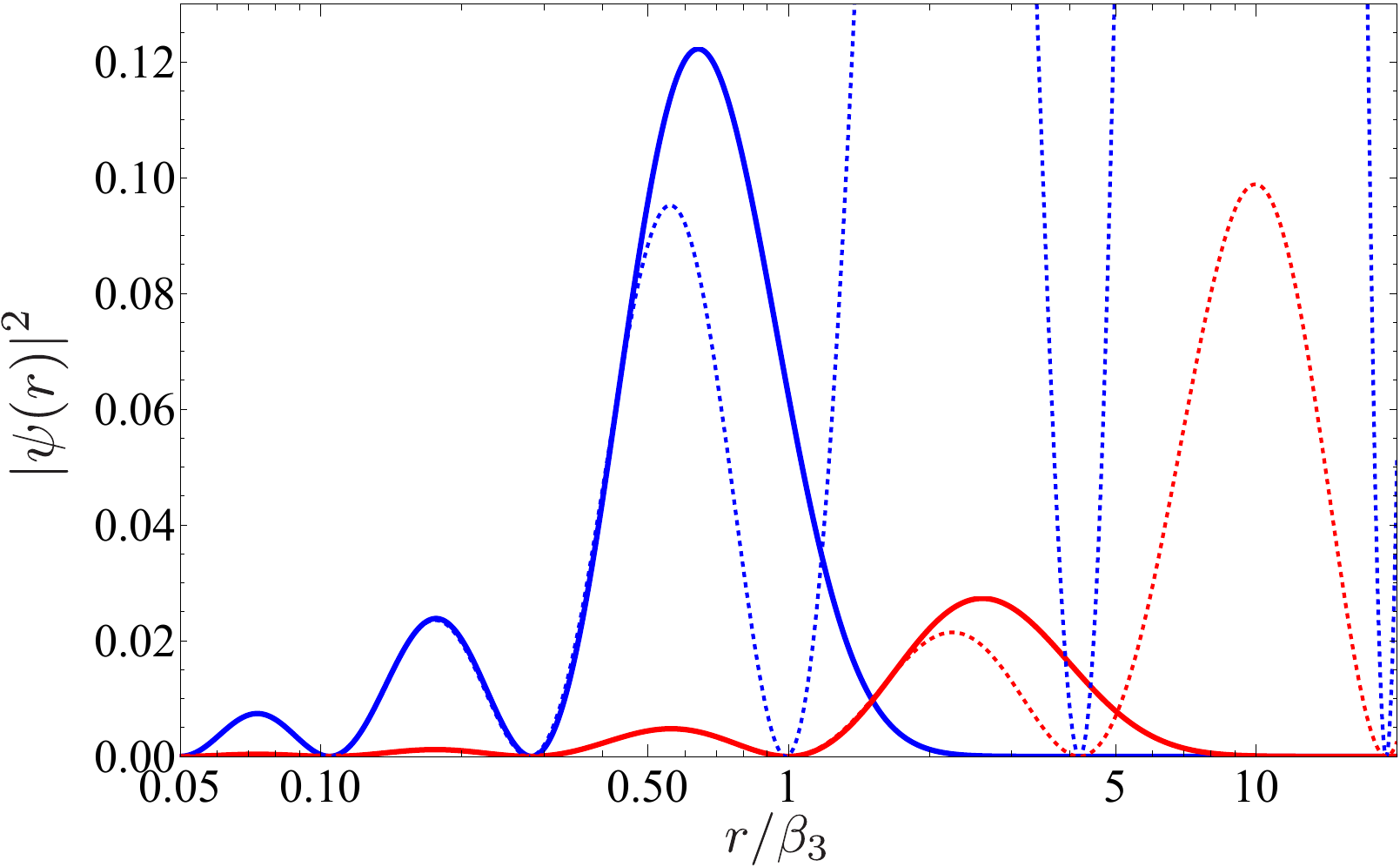} 
\caption{Efimov states' wave function for an attractive $1/r^3$ potential, obtained by solving \equref{onedimSchEq} with $M/m = 27.7520...$, $R_\mathrm{min} / \beta_3 = 0.05$, and $R_\mathrm{max} / \beta_3 = 20.0$. Blue (red) solid curves are the ground and first-excited Efimov states, respectively. Dashed curves denote analytical solutions for $r \ll 1/\sqrt{M |E|}$ in Eqs.~(\ref{eq:zeroenergy_wavefunction_attractive}) and (\ref{eq:QDTWF_fcKcgc}) with $K^c= f^c (R_{\mathrm{min}})/g^c (R_{\mathrm{min}})$.}
\label{fig:AttractWavefunc}
\end{figure}

In \figref{attractive_1DUnitary_BE}(a), we show the numerical solution of \equref{onedimSchEq} with an attractive $1/r^3$ potential $C_3>0$ at the unitary limit $1/a^{(\mathrm{HL})}=0$. They agree excellently with the analytical expression (solid curves) in Eqs.~(\ref{eq:EnergyAttDipole}) and (\ref{eq:kc}) for most regions. In contrast to the repulsive case, the energy spectra depend on $\beta_3/R_{\mathrm{min}}$ through the quantum defect parameter $K^c$, as described by the universal formula in \equref{EnergyAttDipole}. This is consistent with behavior of the wave function shown in \figref{AttractWavefunc}: both the ground (blue) and first-excited (red) Efimov states' wave functions exhibit rapid oscillations and penetrates into short distance in \figref{AttractWavefunc}, in stark contrast to the repulsive case of \figref{RepulWavefunc} where the wave function decays exponentially without such oscillations. The numerical results (solid curves) in \figref{AttractWavefunc} are in agreement with the analytical expression for $r \ll 1/\sqrt{M|E|}$ in \equref{zeroenergy_wavefunction_attractive} (dashed curves) [see also the leading order term of \equref{ShortRangeAttractive}], whereas they disagree at long distances where the condition $r \ll 1/\sqrt{M|E|}$ to derive \equref{zeroenergy_wavefunction_attractive} is not satisfied. 

A discrepancy between the numerical and analytical results for $M \beta_3^2 |E| \lesssim 10^{-2}$ and $\beta_3/R_{\mathrm{min}} \gtrsim 50$ in \figref{attractive_1DUnitary_BE}(a) (upper-right region) is due to the finite-size effect of the trimer becoming similar size as the large-distance boundary. On the other hand, a discrepancy for deeply bound states $M \beta_3^2 |E| \gtrsim 10$ (bottom region) is due to the breakdown of the low-energy condition $M \beta_3^2 |E| \ll 1$ used in deriving \equref{EnergyAttDipole}.

The energy spectra away from the unitary limit $1/a^{(\mathrm{HL})}\neq 0$ are shown in \figref{attractive_1DUnitary_BE}(b). While different $\beta_3/R_{\mathrm{min}}$ generally yield different values of $K^c$ hence different spectra, they are confirmed to collapse almost perfectly onto the same universal curves when the corresponding $K^c$ value (i.e., the three-body short-range phase) is identical, with a small noticeable deviation for the ground Efimov state.

\subsection{Effect of axial harmonic confinement}
In this section, we study the effect of an axially symmetric harmonic confinement according to \equref{q1Dscheq}. By comparing the results of one-dimensional limit $\omega \rightarrow \infty$ in Secs.~\ref{sec:repandatt} and  \ref{sec:attandatt} with those for finite $\omega$ in \equref{q1Dscheq}, we investigate the extent to which our analytical solutions in Secs.~\ref{sec:repandatt} and  \ref{sec:attandatt} are useful for describing the Efimov states of dipolar atoms or polar molecules in a mixed dimension.

\begin{figure}[!t]
	\centering
	\includegraphics[width=1.0\linewidth]{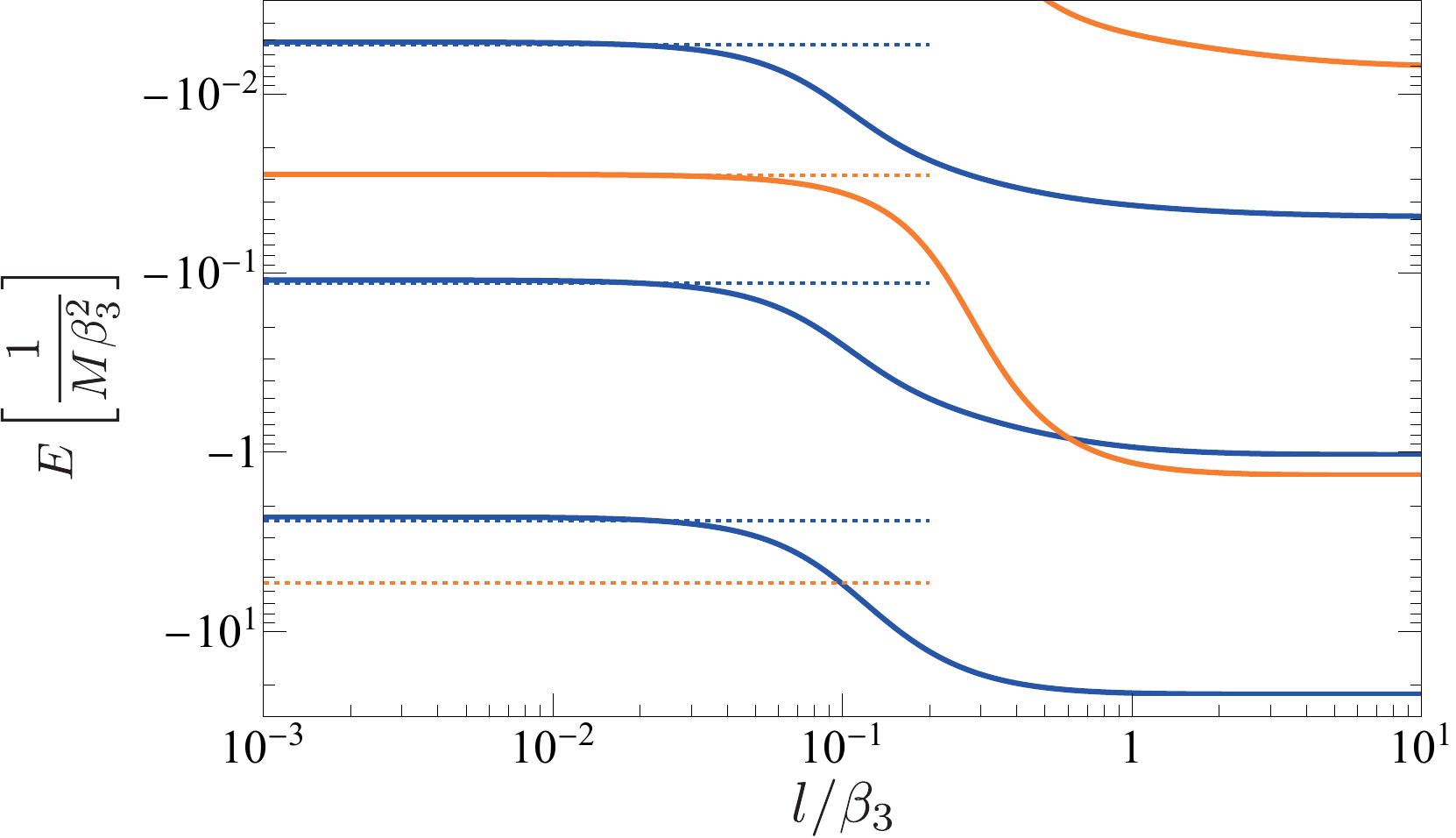} 
\caption{Energy spectra of the Efimov bound states with a repulsive $1/r^3$ potential under an axially symmetric confinement with a harmonic oscillator length $l=\sqrt{\hbar / M \omega}$, calculated with $R_\mathrm{min} / \beta_3 = 0.1$ and $R_\mathrm{max} / R_\mathrm{min} = 1000$. Blue and orange solid curves show the energies obtained by numerically solving \equref{onedimSchEq} for mass ratios corresponding to $^{167}$Er-$^{6}$Li atoms ($M/m = 27.7520...$), and $^{6}$Li atom and $^{23}$Na$^{40}$K molecule ($M/m = 10.4659...$), respectively. The dashed lines denote the analytical result in the one-dimensional limit [\equref{EnergyRepDipole}].}
\label{fig:repulsive_trapUnitary_BE}
\end{figure}

In \figref{repulsive_trapUnitary_BE}, we show the numerical solution of \equref{q1Dscheq} with a repulsive $1/r^3$ potential $C_3<0$ for variable values of harmonic trapping strength $l=\sqrt{\hbar / M \omega}$ at the unitary limit $1/a^{(\mathrm{HL})}=0$. For a tight harmonic confinement $l/\beta_3 \ll 1$, the results agree excellently with the analytical expression in \equref{EnergyRepDipole} (dotted lines), except for the tightly bound state which does not appear for the $^{6}$Li atom-$^{23}$Na$^{40}$K molecule (orange), whose energy $M \beta_3^2 |E| \gtrsim 10$ is too large to apply the low-energy analytical expression in \equref{EnergyRepDipole}. As the trapping strength is weakened, the binding energies increase because of the reduction of the effective dipole repulsion, as the heavy particles can explore three-dimensional configurations and thereby avoid the $1/r^3$ repulsion more efficiently~\cite{PhysRevLett.99.140406, PhysRevA.81.063616}. The discrete-scale invariance with the same scale factor $e^{2\pi/|s|}$ persists even for $l/\beta_3 \gtrsim 1$, as it originates from the common long-range behavior of the potential $V(r)= - (|s|^2 +1/4)/Mr^2$ at large distance $r\gg \beta_3, l$.

\begin{figure}[!t]
	\centering
	\includegraphics[width=1.0\linewidth]{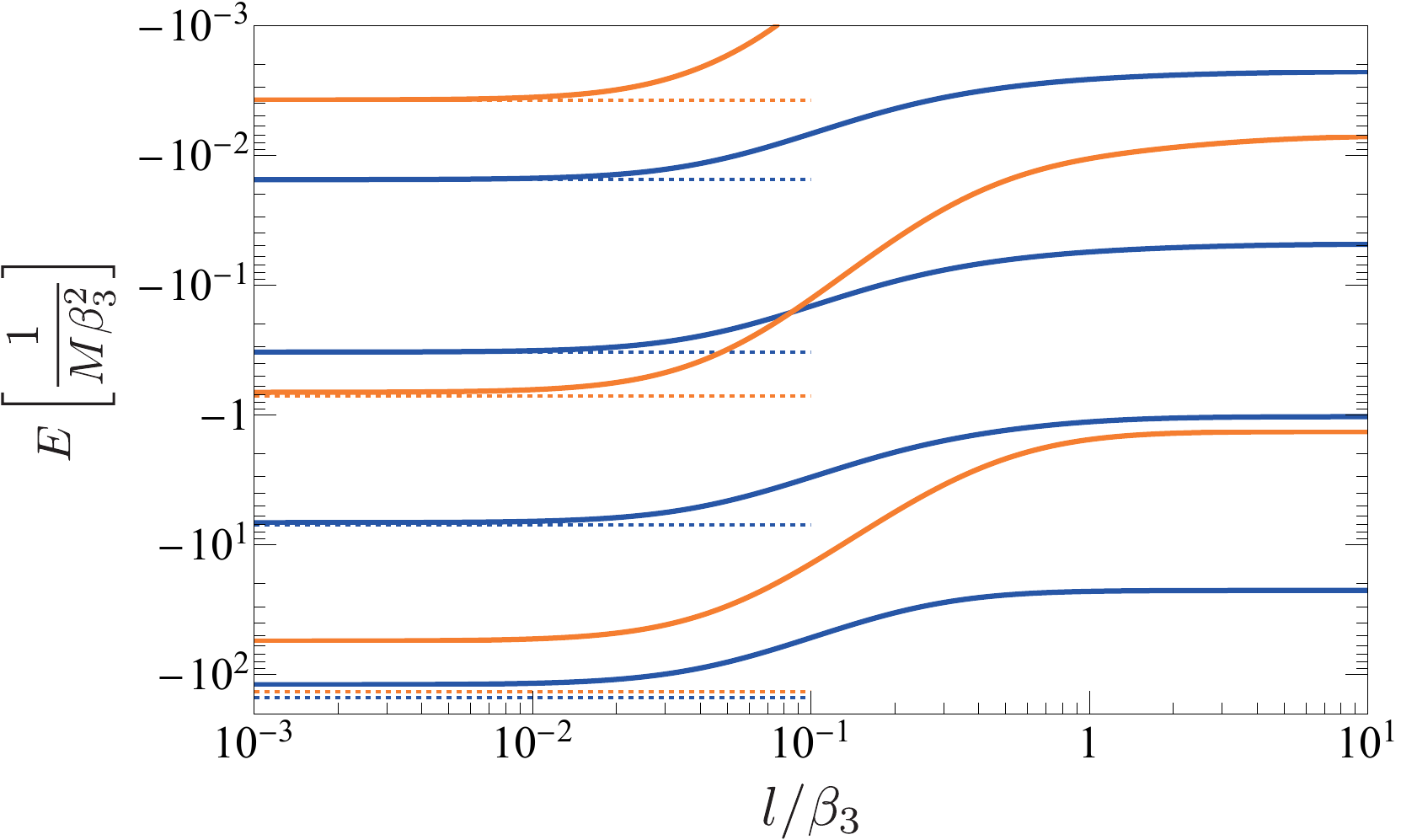} 
\caption{Energy spectra of the Efimov bound states with an attractive $1/r^3$ potential under an axially symmetric confinement, calculated with $R_\mathrm{min} / \beta_3 = 0.1$ and $R_\mathrm{max} / R_\mathrm{min} = 1000$. Dashed lines denote the analytical result in the one-dimensional limit [Eqs.~(\ref{eq:EnergyAttDipole}) and (\ref{eq:kc})].}
\label{fig:attractive_trapUnitary_BE}
\end{figure}

In \figref{attractive_trapUnitary_BE}, we show the results with an attractive $1/r^3$ potential $C_3>0$ at the unitary limit $1/a^{(\mathrm{HL})}=0$. In the tight-trap limit $l/\beta_3 \ll 1$, the results agree excellently with the analytical expression in Eqs.~(\ref{eq:EnergyAttDipole}) and (\ref{eq:kc}) (dotted lines). Discrepancies for the tightly bound states of $M \beta_3^2 |E| \gtrsim 10^2$ in \figref{attractive_trapUnitary_BE} arises from the breakdown of the low-energy approximation. As the trapping strength is weakened, the binding energies decrease because of the reduction of the effective dipole attraction, as the heavy particles take three-dimensional configurations with less $1/r^3$ attractions~\cite{PhysRevLett.99.140406, PhysRevA.81.063616}. Similar to the repulsive case, the discrete-scale invariance with the same scale factor $e^{2\pi/|s|}$ persists for $l/\beta_3 \gtrsim 1$.

While it is challenging to realize the one-dimensional limit $l/\beta_3 \rightarrow 0$ with dipolar atoms, the condition $l/\beta_3 \ll 1$ is feasible using polar molecules. Indeed, the experimental parameters for the dipolar $^{166}$Er atom~\cite{Chomaz_2023} and the $^{23}$Na$^{40}$K polar molecule~\cite{shi2025universal} are estimated, using \equref{C3_theta_rel} and $\theta=0$ with a typical trapping frequency $\omega/2\pi=20$ kHz~\cite{PhysRevX.8.021030,science.abb4928}, as $l/\beta_3= 2.6$ and $l/\beta_3= 0.08$, respectively. As explained in \secref{model} [see \equref{q1deq_vpotang}], $\beta_3$ can be controlled from attractive ($0\le \theta < 54.7^{\circ}$) to repulsive ($\theta > 54.7^{\circ} $) values by varying the orientation of a static external field. The one-dimensional limits for the repulsive and attractive systems, as well as the crossover between them, are thus within reach of current experimental platforms, such as polar molecules confined in a tightly cigar-shaped trap, or possibly Rydberg atoms~\cite{PRXQuantum.3.030339}, which can possess much larger dipole moments.

We note that the quasi-one-dimensional approach used in our work is only applicable to low-energy states. Indeed, it has been demonstrated for two confined dipoles that the quasi-one-dimensional approximation can break down for a distance between two dipoles $r\lesssim l$~\cite{PhysRevA.89.023604,zdziennicki2025}. This implies that the results presented in Figs~\ref{fig:repulsive_trapUnitary_BE} and \ref{fig:attractive_trapUnitary_BE} are not be reliable in the right-bottom region $E \lesssim -1/M l^2$. A three-dimensional treatment, which properly includes excitations into higher transversal modes and resonances induced by them~\cite{PhysRevLett.91.163201}, is required to accurately describe those trimers with relatively large binding energies.

\section{\label{sec:Concl}Conclusion}
We have studied the Schr\"{o}dinger equation with $1/r^3$ and attractive $1/r^2$ potentials and obtained their analytical solutions. By applying the quantum defect theory with complex angular momentum, we have derived the analytical expressions of the binding energies and wave functions in the low-energy regime for both attractive and repulsive $1/r^3$ potentials, and verified their validity through comparison with numerical solutions. Our work is relevant not only from a mathematical perspective in providing analytical solutions of the Schr\"{o}dinger equation~\cite{PhysRevA.78.012702, PhysRevA.59.2778, PhysRevA.64.010701, PhysRevA.72.042719, BoGao2004, PhysRevA.64.022717, OiEndo2024, oi2025universal, PhysRevResearch.7.023162,du2024analytical}, but also for understanding the physical properties of the Efimov states with long-range interactions. Indeed, the Schr\"{o}dinger equation with $1/r^3$ and attractive $1/r^2$ potentials naturally arises as the Born-Oppenheimer equation of a three-body system composed of a two heavy particles with large permanent dipole moments confined in a quasi-one-dimensional trap interacting resonantly with an unconfined light particle. Using the analytical solution, we find that the three-body parameter of the mixed-dimensional Efimov states is universally determined by the dipole length for a repulsive system, whereas for an attractive system it explicitly depends on the short-range phase encoded in the quantum defect parameter. 

Such mixed-dimensional Efimov states can be realized experimentally using dipolar cold atoms or polar molecules confined in a tight cigar-shaped trap interacting with another atomic species via a Feshbach resonance~\cite{ErLiFR1,ErLiFR2,DyLifeshbach,CrLiFR,yang2022creation, doi:10.1126/science.aau5322, chen2023field}. By investigating the effect of finite transverse confinement, we find that the quasi-one-dimensional analytical solutions are well suited to polar molecules, while the tight-confinement condition is challenging to realize for dipolar atoms. Our findings can thus be tested in experiments with a mixture of polar molecule and a light cold atom by observing enhancements of the three-body loss rate induced by the Efimov states~\cite{kraemer2006evidence, Zaccanti2009, PhysRevLett.103.163202,doi:10.1126/science.1182840, PhysRevLett.113.240402, PhysRevLett.112.250404}, under conditions where the two-body losses are efficiently suppressed, for example, by the microwave shielding~\cite{matsuda2020resonant,anderegg2021observation}. They can also be observed through the RF association~\cite{LompeScience11,PhysRevLett.106.143201,PhysRevLett.108.210406}, or by the coherent oscillation~\cite{PhysRevLett.122.200402}. More broadly, our analytical solutions provide a useful step toward understanding three-body correlations in strongly correlated quantum many-body systems~\cite{CRPHYS3BC,PhysRevLett.106.153005,science.aai8195}.


\begin{acknowledgments}
 We thank Peng Zhang and Wenbin Zhong for fruitful discussions. This work was supported by JSPS KAKENHI Grant Numbers JP22K03492, JP23H01174, and JP25K00217. SE acknowledges support from Matsuo Foundation. KO acknowledges support from Graduate Program on Physics for the Universe (GP-PU) of Tohoku University.
\end{acknowledgments}


\textit{Data availability.} The data that support the findings of this article are openly available \cite{ohishi_data}.

\appendix

\section{\label{app:QDTsolution}Analytical solution with $1/r^3$ and attractive $1/r^2$ potentials}
In this Appendix, we present analytical solutions of \equref{onedimSchEq} for both repulsive and attractive $1/r^3$ potentials. Our approach is based on the Bessel series expansion supplemented with the quantum defect theory as developed in Refs~\cite{PhysRevA.64.022717,PhysRevA.78.012702,PhysRevA.64.010701,PhysRevA.59.2778}, generalized to the case of an attractive $1/r^2$ potential.

\subsection{\label{app:repulsive_QDT}Repulsive case}

For a repulsive $1/r^3$ potential, \equref{onedimSchEq} can be solved analytically by generalizing the solutions in Ref.~\cite{PhysRevA.59.2778} to complex angular momenta as in Ref.~\cite{OiEndo2024}. Namely, for a repulsive case, two independent special solutions 
of \equref{onedimSchEq} are defined as
\begin{equation}
    \begin{split}
    f^c(r)&=\frac{1}{\sin(2\pi\nu)}\left[\frac{\xi(r)}{F(-\nu)}-\frac{\eta(r)}{F(\nu)}\right], \\
    g^c(r)&=-\left[\frac{\xi(r)}{F(-\nu)}+\frac{\eta(r)}{F(\nu)}\right].
    \end{split}
    \label{eq:repulsive_AMI_solution_app}
\end{equation}
Here, $\xi(r)$ and $\eta(r)$ are expressed by expansions of the Bessel function as
\begin{align}
\label{eq:defxieta}
    \xi(r)=\sqrt{r}\sum_{m=-\infty}^{\infty}b_mJ_{\nu+m}(x), \\
    \eta(r)=\sqrt{r}\sum_{m=-\infty}^{\infty}b_mJ_{-\nu-m}(x),
\end{align}
where $x=\sqrt{2\mu E}r$. $b_n,$ $F(\nu)$ are defined in a similar manner to those in Ref.~\cite{PhysRevA.59.2778} as
\begin{gather}
    b_j \equiv \Delta^j\frac{\Gamma(\nu) \Gamma(\nu - \nu_0 + 1) \Gamma(\nu + \nu_0 + 1)}{\Gamma(\nu + j)\Gamma(\nu - \nu_0 + j + 1)\Gamma(\nu + \nu_0 + j + 1)} c_j(\nu), \label{eq:def_bj_pj_repulsive}\\
    b_{-j} \equiv \Delta^j\frac{\Gamma(\nu - j + 1)\Gamma(\nu - \nu_0 - j)\Gamma(\nu + \nu_0 - j)}{\Gamma(\nu + 1) \Gamma(\nu - \nu_0) \Gamma(\nu + \nu_0)} c_j(-\nu), \label{eq:def_bj_mj_repulsive}\\
    c_j(\nu) \equiv b_0 Q(\nu) Q(\nu + 1) \cdots Q(\nu + j -1), \label{eq:def_cj}\\
    Q(\nu) \equiv \frac{1}{1 - \Delta^2 \dfrac{Q(\nu+1)}{(\nu+1) [(\nu+1)^2 - \nu_0^2] (\nu+2) [(\nu+2)^2 - \nu_0^2]}} , \label{eq:def_Qnu}\\
    F(\nu) \equiv |\Delta|^{-\nu} \frac{\Gamma(1+\nu_0+\nu)\Gamma(1-\nu_0+\nu)}{\Gamma(1-\nu)} C(\nu) \label{eq:def_Fnu}, \\
    C(\nu) \equiv \lim_{n\rightarrow\infty} c_n \label{eq:def_capital_cnu},
\end{gather}
where  $\Delta=\sqrt{2\mu E}\beta_3/2$. In Eqs.~(\ref{eq:def_bj_pj_repulsive}) to (\ref{eq:def_cj}), $j$ is a positive integer, and the wavefunction is normalized so that $b_0=1$. $\nu$ is a complex number determined by the transcendental equations~\cite{PhysRevA.59.2778}
\begin{equation}
    (\nu^2 - \nu_0^2) - \frac{\Delta^2}{\nu} [\bar{Q}(\nu) - \bar{Q}(-\nu)] = 0,
\end{equation}
where
\begin{equation}
    \bar{Q}(\nu) \equiv \{ (\nu + 1) [(\nu+1)^2 - \nu_0^2] \}^{-1} Q(\nu).
\end{equation}
Here, $\nu_0=s = i |s|$ is pure imaginary.
By employing the asymptotic form of the Bessel function, we obtain the asymptotic forms of \equref{repulsive_AMI_solution_app} at short distance $r\ll \beta_3$ up to next-to-next-to-leading order as
\begin{equation}
    \begin{split}
    &f^c(r)\simeq\sqrt{\frac{r}{\pi}}\left(\frac{r}{\beta_3} \right)^{\frac{1}{4}}\exp\left(-2\sqrt{\frac{\beta_3}{r}}\right)\\
    &\left[ 1 - \sqrt{\frac{r}{\beta_3}} \left( |s|^2 + \frac{1}{16} \right)  + \frac{r}{2\beta_3} \left( |s|^2+ \frac{9}{16} \right) \left( |s|^2 + \frac{1}{16} \right) \right], \\
    &g^c(r)\simeq-\sqrt{\frac{r}{\pi}}\left(\frac{r}{\beta_3} \right)^{\frac{1}{4}}\exp\left(2\sqrt{\frac{\beta_3}{r}}\right)\\
    &\left[ 1 + \sqrt{\frac{r}{\beta_3}} \left( |s|^2 + \frac{1}{16} \right)  +  \frac{r}{2\beta_3} \left( |s|^2+ \frac{9}{16} \right) \left( |s|^2 + \frac{1}{16} \right) \right]. 
    \end{split}
    \label{eq:ShortRangeRepulsive}
\end{equation}
On the other hand, the long-range asymptotic forms $r\gg \beta_3$ are obtained as 
\begin{equation}
    \begin{split}
    f^c(r)\simeq\dfrac{1}{\sqrt{2\pi\kappa}}\left[ W_{f-}e^{\kappa r}-W_{f+}(2e^{-\kappa r}) \right],\\
    g^c(r)\simeq\dfrac{1}{\sqrt{2\pi\kappa}}\left[ W_{g-}e^{\kappa r}-W_{g+}(2e^{-\kappa r}) \right],
    \end{split}
    \label{eq:LongRangeAsymptotic}
\end{equation}
where $\kappa =\sqrt{2\mu |E|}$, and $W_{f\pm}$ and $W_{g\pm}$ are given as Eqs.~(32) and (35) in Ref.~\cite{PhysRevA.59.2778}. Notably, the leading-order term in \equref{ShortRangeRepulsive} is independent of energy and $s$, which is an essential requirement for the $s$-insensitive (i.e. angular-momentum-insensitive) quantum defect theory employed in our work~\cite{PhysRevA.64.010701}.

A general solution for \equref{onedimSchEq} is expressed using the two independent special solutions as \equref{QDTWF_fcKcgc}. Therefore, a boundary condition for bound states $\psi(r=\infty)=0$ is rewritten as
\begin{equation}
\label{eq:B.C.KcAndWfmWgm}
    K^c=\frac{W_{f-}}{W_{g-}},
\end{equation}
where
\begin{equation}
\label{eq:WfmOverWgm_repulsive}
    \frac{W_{f-}}{W_{g-}}=\dfrac{1}{\sin[2\pi(\nu-\nu_0)]}\dfrac{1-\mathcal{M}}{1+\mathcal{M}}
\end{equation}
with 
\begin{gather}
    \mathcal{M} \equiv G(-\nu)/G(\nu), 
    \label{eq:M_def}
    \\
    G(\nu) \equiv |\Delta|^{-\nu} \frac{\Gamma(1+\nu_0+\nu)\Gamma(1-\nu_0+\nu)}{\Gamma(1-\nu)} C(\nu).
    \label{eq:G_def}
\end{gather}
By performing a low-energy expansion, the binding energy can be obtained analytically: at low energy, $\nu$ and $C(\nu)$ asymptote as 
\begin{align}
\label{eq:nuLowEnergy}
    \nu &\cong \nu_0, \\
    C(\nu) &\cong 1.
    \label{eq:CnuLowEnergy}
\end{align}
Therefore, from the definition of $G(\nu)$, $\mathcal{M}$ asymptotes as
\begin{equation}
    \begin{split}
        \mathcal{M} &= |\Delta|^{2 \nu} \frac{\Gamma(1 + \nu_0 - \nu) \Gamma(1 - \nu_0 - \nu)}{\Gamma(1 + \nu)} \\& \times\frac{\Gamma(1-\nu)}{\Gamma(1 + \nu_0 + \nu)\Gamma(1 - \nu_0 + \nu)}\\
        &\approx |\Delta|^{2\nu_0} \frac{\Gamma(1-2\nu_0)}{\Gamma(1+\nu_0)} \frac{\Gamma(1-\nu_0)}{\Gamma(1+2\nu_0)}.
        \label{eq:Mel_1}
    \end{split}
\end{equation}
In the absence of the Efimov effect, $\nu_0$ is real, as for a two-body problem with an angular momentum $\ell$ under $1/r^3$ potential $\nu_0=2\ell+1$~\cite{PhysRevA.59.2778,PhysRevA.64.022717}.
With the Efimov attraction, on the other hand, $\nu_0$ becomes pure imaginary $\nu_0 = i|s|$. Therefore, $\mathcal{M}$ at low energy is obtained as
\begin{equation}
    \begin{split}
        \mathcal{M} &\cong |\Delta|^{2i|s|} \frac{-2\pi^2}{\mathrm{sinh}(\pi |s|)\mathrm{sinh}(2\pi |s|)} \frac{1}{[\Gamma(i|s|) \Gamma(1+2i|s|)]^2} \\
        &=- |\Delta|^{2i|s|} e^{-2i\xi} = -e^{i \{ 2|s| \ln|\Delta| - 2\xi \}},
    \label{eq:Mel_3}
\end{split}
\end{equation}
where
\begin{equation}
    \xi \equiv \mathrm{arg} [\Gamma(i|s|)\Gamma(1+2i|s|)].
    \label{eq:xiDef}
\end{equation}

As shown in \equref{ShortRangeRepulsive}, $f^c$ decreases exponentially as $r\rightarrow0$ while $g^c$ increases exponentially, corresponding to a regular and irregular solution, respectively. If the length scale associated with the short-range physics $R_{\mathrm{min}}$ is small $R_{\mathrm{min}} \ll \beta_3$, as is the case for cold atoms and molecules, the wave function should be solely described by the regular solution $f^c$. In other words, in the short-range region $r \ll \beta_3$, the potential is dominated by the repulsive $1/r^3$ interaction. Therefore, the wave function should be suppressed in the short-range region, which suggests
\begin{equation}
\begin{split}
    K^c &= \frac{1}{\mathrm{sin}[2\pi (\nu - \nu_0)]}\frac{1-\mathcal{M}}{1 + \mathcal{M}}\\
    &\cong0 \ \ \ \ \text{ $\Delta \rightarrow 0$}.
    \label{eq:KcconditionofB.S.Repulsive}
\end{split}
\end{equation}
Namely,
\begin{equation}
    \mathcal{M} \cong 1.
\end{equation}
Substituting \equref{Mel_1} into this, we find 
\begin{equation}
   |s|\ln |\Delta| - \xi = \left(-n + \frac{1}{2}\right) \pi,
    \label{eq:K0tan}
\end{equation}
from which we obtain the binding energy for a repulsive potential at low energy $|E|\ll1/2\mu  \beta_3^{~2}$ as
\begin{equation}
    |\Delta| = \mathrm{exp}\left\{ \frac{1}{|s|} \left(\left(-n+\frac{1}{2}\right)\pi + \xi\right) \right\}.
    \label{eq:KcDeltaRepulsive}
\end{equation}

\subsection{\label{app:attractive_QDT}Attractive case}

The analytical solution of \equref{onedimSchEq} for an attractive potential can also be obtained in a similar manner. Namely, the two independent solutions are defined as

\begin{equation}
    \begin{split}
    f^c(r)&=\frac{1}{2\cos(\pi\nu)}\left[\frac{\xi(r)}{F(-\nu)}+\frac{\eta(r)}{F(\nu)}\right], \\
    g^c(r)&=\frac{1}{2\sin(\pi\nu)}\left[\frac{\xi(r)}{F(-\nu)}-\frac{\eta(r)}{F(\nu)}\right].
    \end{split}
    \label{eq:attractive_AMI_solution_app}
\end{equation}
Note that the definition of $c_n,Q(\nu),F(\nu)$ and the transcendental equation which determines $\nu$ are the same as those in the repulsive case. On the other hand, the equations between $b_j$ and $c_j(\nu)$ are  different from those in the repulsive case, and are given as~\cite{PhysRevA.64.022717}, 
\begin{gather}
    b_j \equiv (-\Delta)^j\frac{\Gamma(\nu) \Gamma(\nu - \nu_0 + 1) \Gamma(\nu + \nu_0 + 1)}{\Gamma(\nu + j)\Gamma(\nu - \nu_0 + j + 1)\Gamma(\nu + \nu_0 + j + 1)} c_j(\nu), \label{eq:def_bj_pj_attractive}\\
    b_{-j} \equiv (-\Delta)^j\frac{\Gamma(\nu - j + 1)\Gamma(\nu - \nu_0 - j)\Gamma(\nu + \nu_0 - j)}{\Gamma(\nu + 1) \Gamma(\nu - \nu_0) \Gamma(\nu + \nu_0)} c_j(-\nu), \label{eq:def_bj_mj_attractive}
\end{gather}
With this slight difference in mind, a general solution for \equref{onedimSchEq} is expressed by a linear combination of $f^c$ and $g^c$ as in \equref{QDTWF_fcKcgc}.

The short-range asymptotic forms of \equref{attractive_AMI_solution_app} for $r\ll \beta_3$ up to next-to-next-to-leading order can be obtained as
\begin{equation}
    \begin{split}
    f^c(r)\simeq & \sqrt{\frac{r}{\pi}}\left(\frac{r}{\beta_3}\right)^{\frac{1}{4}} \left[ 1 - \frac{r}{2\beta_3} \left( |s|^2+\frac{9}{16} \right) \left( |s|^2+\frac{1}{16} \right) \right]\cos\Phi_r \\
    & +\sqrt{\frac{r}{\pi}} \left(\frac{r}{\beta_3}\right)^{\frac{3}{4}} \left( |s|^2+ \frac{1}{16} \right) \sin\Phi_r+ O\left( \left( \frac{r}{\beta_3} \right)^{\frac{9}{4}} \right) , 
    \\
    g^c(r) \simeq & -\sqrt{\frac{r}{\pi}}\left(\frac{r}{\beta_3}\right)^{\frac{1}{4}} \left[ 1 - \frac{r}{2\beta_3} \left( |s|^2+\frac{9}{16} \right) \left( |s|^2+\frac{1}{16} \right) \right] \sin\Phi_r\\
     & +\sqrt{\frac{r}{\pi}} \left(\frac{r}{\beta_3}\right)^{\frac{3}{4}}  \left( |s|^2+ \frac{1}{16} \right) \cos\Phi_r+ O\left( \left( \frac{r}{\beta_3} \right)^{\frac{9}{4}} \right). 
    \label{eq:ShortRangeAttractive}
    \end{split}
\end{equation}
where $\Phi_r=2\sqrt{\beta_3/r}-\pi/4$. The long-range asymptotic form in the attractive case is given by the same expression as in \equref{LongRangeAsymptotic} with $W_{f\pm}$ and $W_{g\pm}$ given as
\begin{equation}
    \begin{split}
    W_{f-}&=\dfrac{1}{2 \cos \left( \nu \pi \right) } \dfrac{1}{G \left( -\nu \right) } \left( 1+\mathcal{M}\right)\left( \sum ^{\infty }_{m=-\infty }i^{m}b_{m}\right), \\
    W_{f+}&=\dfrac{\tan \left( \nu \pi \right)}{2} \dfrac{1}{G \left( -\nu \right) } \left( 1-\mathcal{M}\right)\left( \sum ^{\infty }_{m=-\infty }(-i)^{m}b_{m}\right), \\
    W_{g-}&=\dfrac{1}{2 \sin \left( \nu \pi \right) } \dfrac{1}{G \left( -\nu \right) } \left( 1-\mathcal{M}\right)\left( \sum ^{\infty }_{m=-\infty }i^{m}b_{m}\right), \\
    W_{g+}&=\dfrac{1}{2} \dfrac{1}{G \left( -\nu \right) } \left( 1+\mathcal{M}\right)\left( \sum ^{\infty }_{m=-\infty }(-i)^{m}b_{m}\right),
    \label{eq:Wfg}
    \end{split}
\end{equation}
where the definitions of $G(\nu)$ and $\mathcal{M}$ are the same as those in the repulsive case.

The binding energy can be obtained by introducing the boundary condition $\psi(r=\infty)=0$ as
\begin{equation}
    \label{eq:WfmOverWgm_attractive}
    K^c = \frac{W_{f-}}{W_{g-}}=\tan(\nu \pi)\frac{1+\mathcal{M}}{1-\mathcal{M}}.
\end{equation}
For an attractive potential, the right-hand side of \equref{WfmOverWgm_attractive} at low energy regime can be approximated using Eqs.~(\ref{eq:nuLowEnergy}), (\ref{eq:CnuLowEnergy}) and (\ref{eq:Mel_1}) (which are also valid for the attractive case) as
\begin{equation}
\label{eq:WfmWgmLowenergyAttractive}
    \frac{W_{f-}}{W_{g-}} \cong \tanh(\pi |s|)\tan \left[\frac{|s|}{2}\ln|\Delta|-\xi\right].
\end{equation}
Substituting this into the bound-state condition \equref{WfmOverWgm_attractive}, the binding energy at low-energy regime $|E|\ll1/2\mu \beta_3^{~2}$ is obtained as
\begin{equation}
    |\Delta|= \mathrm{exp} \left( \frac{1}{|s|} \left[\mathrm{arctan}\left( \frac{K^c}{\mathrm{tanh}(\pi |s|)} \right) +\xi \right] \right) \times e^{-\frac{n\pi}{|s|}}.
    \label{eq:KcDeltaAttractive}
\end{equation}
For an attractive case, the three-body binding energy depends explicitly on the QDT parameter $K^c$. 

To determine the QDT parameter $K^c$, a boundary condition at short range region is needed. In this study, we adopt the hard-wall boundary condition $\psi(r=R_{\mathrm{min}})=0$, from which $K^c$ is obtained as
\begin{equation}
\begin{split}
    K^c &= -\mathrm{tanh} (\pi |s|) \frac{\mathrm{Re}\left[ J_{2i|s|} \left(2 \sqrt{\frac{\beta_3}{R_{\mathrm{min}}}}\right)\right]}{\mathrm{Im}\left[J_{2i|s|}\left(2 \sqrt{\frac{\beta_3}{R_{\mathrm{min}}}}\right)\right]} \\
    &\cong -\frac{1}{\mathrm{tan}\left[2\sqrt{\frac{\beta_3}{R_\mathrm{min}}} - \frac{\pi}{4}\right]}.
    \end{split}
    \label{eq:JBesselKc}
\end{equation}
 In the last line, $R_{\mathrm{min}} \ll \beta_3$ is assumed to obtain the asymptotic form of  $K^c$. Notably, $K^c$ for small $R_{\mathrm{min}}$ is independent of energy nor $s$, which is consistent with the $s$-insensitive (i.e. angular-momentum-insensitive) quantum defect theory used in our study~\cite{PhysRevA.64.010701}.

\section{\label{app:zeroenergy}Zero-energy solution}

To clarify the relations between the zero-energy solutions Eqs.~(\ref{eq:fcgc_shortI_rep}) and (\ref{eq:zeroenergy_wavefunction_attractive}) and the finite-energy solutions Eqs.~(\ref{eq:repulsive_AMI_solution_app}) and (\ref{eq:attractive_AMI_solution_app}), we present here the derivation of the zero-energy solutions by taking the zero-energy limit of the finite-energy solutions.

\subsection{\label{app:zeroenergy_repulsive}Repulsive case}
First, we derive the zero-energy solutions \equref{fcgc_shortI_rep} in the repulsive case. From \equref{defxieta}, $\xi(r)$ in \equref{repulsive_AMI_solution_app} is reformulated as
\begin{equation}
    \begin{split}
        \xi(r) &= \sqrt{r} \sum_{m=-\infty}^{m=\infty}b_m J_{\nu+m}(x) \\
        &=\sqrt{r} \sum_{m=-\infty}^{m=\infty} \sum_{s=0}^{s=\infty} b_m \frac{(-1)^s}{s! \Gamma(s+\nu+m+1)} \left(\frac{x}{2}\right)^{2s+\nu+m} \\
        &=\sqrt{r} \sum_{m=-\infty}^{m=\infty} \sum_{s=0}^{s=\infty} b_{-m-2s} \frac{(-1)^s}{s! \Gamma(\nu-m-s+1)} \left(\frac{x}{2}\right)^{\nu-m}.
    \end{split}
\end{equation}
In the last line, $m$ is replaced by $-m-2s$. Introducing $y \equiv 2 \sqrt{\beta_3/r}$, (i.e., $x=8\Delta /y^2$), we obtain
\begin{equation}
    \begin{split}
       \xi(r) &=\sqrt{r} \sum_{m=-\infty}^{m=\infty} \sum_{s=0}^{s=\infty} b_{-m-2s} \frac{(-1)^s}{s! \Gamma(\nu-m-s+1)} \left(\frac{4\Delta}{ y^2}\right)^{\nu-m} \\
       &=\sqrt{r} \left(\frac{y}{2}\right)^{-2\nu}\sum_{m=-\infty}^{m=\infty} p_m \left(\frac{y}{2}\right)^{2m},
    \end{split}
    \label{eq:xi_with_pm_repulsive}
\end{equation}
where $p_m$ is defined as Eq.~(A15) in Ref.~\cite{PhysRevA.59.2778}
\begin{equation}
    p_m \equiv \Delta^\nu \sum_{s=0}^{s=\infty} \frac{(-1)^s \Delta^{2s}}{s! \Gamma(\nu-m-s+1)} \Delta^{-(m+2s)} b_{-m-2s}.
    \label{eq:def_pm}
\end{equation}
At low-energy $|\Delta | \ll 1 $, large $m$ terms in $\xi(r)$ are dominant and $p_m$ can be replaced with $\lim_{m\rightarrow\infty} p_m$. As shown in Eq.~(A16) in Ref.~\cite{PhysRevA.59.2778}, the limit of large $m$ of $p_m$ is given as
\begin{equation}
    \lim_{m\rightarrow\infty} p_m = F(-\nu) \frac{1}{m!\Gamma(-2\nu+m+1)}.
    \label{eq:largemlimit_pm_repulsive}
\end{equation}
Replacing $p_m$ in \equref{xi_with_pm_repulsive} with $\lim_{m\rightarrow\infty} p_m$, we obtain the zero-energy limit of $\xi(r)$ as
\begin{equation}
    \begin{split}
        \lim_{E\rightarrow 0} \xi(r) &= \sqrt{r} F(-\nu_0) \sum_{m=-\infty}^{m=\infty}\frac{1}{m!\Gamma(-2\nu_0+m+1)} \left(\frac{y}{2}\right)^{2m-2\nu_0} \\
        &=\sqrt{r} F(-\nu_0) I_{-2\nu_0}\left(2\sqrt{\frac{\beta_3}{r}}\right),
    \end{split}
\end{equation}
where the low-energy property $\nu \simeq \nu_0 = s$ is used in the first line.

Similarly, the zero-energy limit of $\eta(r)$ is obtained as
\begin{equation}
    \lim_{E\rightarrow 0}\eta(r)=\sqrt{r} F(\nu_0) I_{2\nu_0}\left(2\sqrt{\frac{\beta_3}{r}}\right).
\end{equation}
Then, from \equref{repulsive_AMI_solution_app}, we obtain the zero-energy limits of $f^c$ and $g^c$ as
\begin{equation}
    \begin{split}
     \lim_{E\rightarrow 0} f^c(r) &= -\frac{2}{\sinh(2 \pi |s|)} \mathrm{Im}\left[\sqrt{r} I_{2i|s|}\left(2 \sqrt{\frac{\beta_3}{r}} \right) \right],\\
     \lim_{E\rightarrow 0} g^c(r) &= - 2 \mathrm{Re}\left[\sqrt{r} I_{2i|s|}\left(2 \sqrt{\frac{\beta_3}{r}} \right) \right].
    \end{split}
    \label{eq:app_zeroenergy_wavefunction_repulsive}
\end{equation}

\subsection{\label{app:zeroenergy_attractive}Attractive case}
The zero-energy solution in the attractive case [\equref{zeroenergy_wavefunction_attractive}] is obtained in a similar manner as in the repulsive case. Following the same procedure as \equref{xi_with_pm_repulsive}, $\xi(r)$ is reformulated as
\begin{equation}
    \xi(r)=\sqrt{r} \left(\frac{y}{2}\right)^{-2\nu}\sum_{m=-\infty}^{m=\infty} p_m \left(\frac{y}{2}\right)^{2m},
    \label{eq:xi_with_pm_attractive}
\end{equation}
where $p_m$ is given as \equref{def_pm}. Taking into account the difference of the definition of $b_n$ between the repulsive and attractive cases, $\lim_{m \rightarrow \infty}p_m$ is given as
\begin{equation}
    \lim_{m\rightarrow\infty} p_m = F(-\nu) \frac{(-1)^m}{m!\Gamma(-2\nu+m+1)}.
    \label{eq:largemlimit_pm_attractive}
\end{equation}
Replacing $p_m$ in \equref{xi_with_pm_attractive} with $\lim_{m\rightarrow\infty} p_m$, we obtain the zero-energy limit of $\xi(r)$ as
\begin{equation}
    \begin{split}
        \lim_{E\rightarrow 0} \xi(r) &= \sqrt{r} F(-\nu_0) \sum_{m=-\infty}^{m=\infty}\frac{(-1)^m}{m!\Gamma(-2\nu_0+m+1)} \left(\frac{y}{2}\right)^{2m-2\nu_0} \\
        &=\sqrt{r} F(-\nu_0) J_{-2\nu_0}\left(2\sqrt{\frac{\beta_3}{r}}\right),
    \end{split}
\end{equation}
where the low-energy property $\nu \simeq \nu_0 = s$ is used in the first line.

Similarly, the zero-energy limit of $\eta(r)$ is obtained as
\begin{equation}
    \lim_{E\rightarrow 0}\eta(r)=\sqrt{r} F(\nu_0) J_{2\nu_0}\left(2\sqrt{\frac{\beta_3}{r}}\right).
\end{equation}
Then, from \equref{attractive_AMI_solution_app}, we obtain the zero-energy limits of $f^c$ and $g^c$ as
\begin{equation}
    \begin{split}
    \lim_{E\rightarrow 0}f^c(r) = \frac{1}{\cosh(\pi |s|)} \mathrm{Re}\left[\sqrt{r} J_{2i|s|}\left(2 \sqrt{\frac{\beta_3}{r}} \right) \right],\\
    \lim_{E\rightarrow 0}g^c(r) = -\frac{1}{\sinh(\pi |s|)} \mathrm{Im}\left[\sqrt{r} J_{2i|s|}\left(2 \sqrt{\frac{\beta_3}{r}} \right) \right].
    \end{split}
    \label{eq:app_zeroenergy_wavefunction_attractive}
\end{equation}

\end{document}